\documentclass[aps,orb,amssymb,showpacs,prd]{revtex4-2}
\usepackage{graphicx}
\usepackage{color}
\usepackage{dcolumn}
\usepackage{color}
 \begin{document}

 \def\en{\textsf{e}}
\def\et{\epsilon_t}
\def\bc{\textbf{c}}
\def\bg{\textbf{g}}
\def\bk{\textbf{k}}
\def\a{{\alpha }}
\def\bt{{\beta }}
\def\be{\begin{equation}}
\def\ee#1{\label{#1}\end{equation}}
\def\d{\textsf{d} }
\def\b{\textsf{b} }
\def\bu{\textbf{u}}
\def\bx{\textbf{x}}
\def\bw{\textbf{w}}
\def\bJ{\textbf{J}}
\def\wa{\widetilde\alpha}
\def\wb{\widetilde\beta}
\def\k{\textbf{k}}

\newcommand{\ben}{\begin{eqnarray}}
\newcommand{\een}{\end{eqnarray}}
\def\lb{\label}
\def\no{\nonumber}

\title{Moderately dense granular gas of inelastic rough spheres}

\author{Gilberto M. Kremer}\email{kremer@fisica.ufpr.br}\affiliation{Departamento de F\'{\i}sica, Universidade Federal do Paran\'a, Caixa Postal 19044, 81531-980 Curitiba, Brazil}

\begin{abstract}

  A kinetic theory for moderately dense gases of inelastic and rough spherical molecules is developed from 
the Enskog equation where  a macroscopic state is characterised   by 29 scalar fields which correspond to the moments of the distribution function: mass density, hydrodynamic velocity, pressure tensor, absolute temperature, translational and rotational heat fluxes, hydrodynamic angular velocity and angular velocity flux. The  balance equations for the 29 scalar fields are obtained from a transfer equation derived from the Enskog equation where  the kinetic and potential parts of the new moments of the distribution function and production terms are calculated from Grad's distribution function for the basic fields.
The transition from the 29 field theory to an eight field theory -- with mass density, hydrodynamic velocity, absolute temperature and hydrodynamic angular velocity -- leads to the determination of the transport coefficients of the Navier-Stokes and Fourier laws. The transport coefficients are functions of the normal and  tangential restitution coefficients and of the local equilibrium radial distribution function. The transport coefficients in the limiting case of elastic rough spheres is also determined.
\end{abstract}
 \maketitle

\section{Introduction}\lb{sec1}

The elastic collisions of the molecules of a rarefied gas without external forces bring the gas to an equilibrium state described by a Maxwellian distribution function. For inelastic collisions of the gas molecules  a part of the translational kinetic energy is converted into heat which implies  a temperature decay of the gas due to the lost of mechanical energy. Granular gases refer to gases where inelastic collisions of  the molecules hold. The statistical description of granular gases is a topic of significant research in the literature and among other works we quote the papers \cite{r1,r2,r3,r4,r5,r6,r7,r8,r9,r10,r11,r12,r13,r14,r15,r16,r17,r18,r19,r20,r21}, the books \cite{b1,b2} and the references therein.

The molecules of polyatomic gases are characterised to have internal degrees of freedom associated with rotational and vibrational energies as well as to electronic and nuclei states. The most simple model of a polyatomic molecule which takes into account the rotational energy without the necessity to specify the space orientation of the molecule is the one proposed by Bryan (see \cite{r22,CC,gk0}). In this model the spherical molecules are considered to be perfectly rough, elastic and rigid. During an elastic binary collision the rough and spherical molecules  grip to each other without slipping, so that the relative velocity of the points of contact is reverse at collision.

The development of a kinetic theory of polyatomic gases with rough elastic spherical molecules and the determination of the transport coefficients was first analysed by Pidduck \cite{r22} in 1922. The extension of the theory to moderately dense gases on the basis of  the Enskog kinetic equation -- which is a modification of the Boltzmann equation for moderately dense gases -- was investigated in \cite{r23} and in \cite{gk1} where a 29-field theory based on Grad's moment method was considered.

Grad's moment method was also applied in the development of a kinetic theory for a rarefied and moderately dense granular gases with smooth inelastic spherical molecules (see e.g. \cite{r2,r3,r10,r13,r15,r18,r19}).

The transport coefficients for rarefied granular gases of inelastic and rough spherical molecules were determined by using the hard-sphere model \cite{r16} and more recently from a model of inelastic and  Maxwell particles \cite{r30}. In a recent paper Garz\'o \cite{r21} pointed out that  the results derived in \cite{r16} for the transport coefficients should be extended to moderate densities.

The aim of the present paper is to develop a kinetic theory for moderately dense gases of inelastic and rough spherical molecules from 
the Enskog equation and we follow the  methodology proposed in \cite{gk1} and characterise a macroscopic state of the  moderately dense gas  by 29 scalar fields which correspond to the moments of the distribution function, namely, mass density, hydrodynamic velocity, pressure tensor, absolute temperature, translational and rotational heat fluxes, hydrodynamic angular velocity and angular velocity flux. The corresponding balance equations for the 29 scalar fields are obtained from a transfer equation derived from the Enskog equation. From the knowledge of  Grad's distribution function for the basic fields the kinetic and potential parts of the new moments of the distribution function and production terms are calculated.  From the 29 field theory a transition to an eight field theory -- with mass density hydrodynamic velocity, absolute temperature and hydrodynamic angular velocity -- is performed which leads to the determination of the transport coefficients of the Navier-Stokes and Fourier laws as functions of the normal and  tangential restitution coefficients and of the local equilibrium radial distribution function. 

In the thirteen method of moments of Grad for rarefied monatomic gas  the basic fields are the mass density, hydrodynamic velocity, absolute temperature and their fluxes:  stress tensor and heat flux vector. Here the 29 moments of the distribution function are the mass density, hydrodynamic velocity, absolute temperature, hydrodynamic angular velocity and their fluxes: stress tensor,  translational and rotational heat fluxes,  and angular velocity flux.

 In 1991 Lun \cite{Lun} developed a kinetic theory for granular flow of dense, inelastic and  rough spheres  by considering the mass density, hydrodynamic velocity, translational and rotational temperatures and hydrodynamic angular velocity as basic fields.  The  constitutive terms in this theory were the pressure tensor, the translational and rotational heat fluxes, the angular velocity flux and the  translational and rotational cooling rates. The constitutive equations for the pressure tensor, translational and rotational  heat fluxes were calculated from a Grad's distribution function, which was a function of  the basic fields, the pressure tensor and the translational and rotational heat fluxes.

The present work differs from \cite{Lun} since it is an extension of Grad's 13-moment method to a dense granular gas of inelastic rough spheres by considering a 29-moment method.  The scheme to obtain the constitutive equations of lower order moments by using higher order moment equations is well-known in the literature which we follow here. The difference of the transport coefficients here and those of Lun is that they depend on the cooling rate. In the case of inelastic smooth sphere this methodology matches the transport coefficients obtained from the Chapman-Enskog method (see e.g. \cite{b2,r15,r18,r19}).  Here the field equations for the translational and rotational heat fluxes and  angular velocity flux are coupled  which has an influence on the transport coefficients of the translational and rotational heat fluxes as well as of the angular velocity flux. Furthermore, the constitutive equation for the cooling rate is determined in this work.

The organisation of the paper is the following: in Section \ref{sec2} the dynamics of of binary collision for inelastic rough spherical molecules is introduced and the transfer equation is derived from Enskog equation for moderately dense gases. The determination of the balance equations for the 29 scalar fields is the subject of Section \ref{sec3}, where Grad's distribution function is introduced and the  kinetic and potential parts of the new moments of the distribution function and production terms are calculated. In Section \ref{sec4} the transition  to an eight field theory is analysed, the corresponding constitutive equations are calculated and the new transport coefficients are identified. 

Cartesian notation for tensors is used with angular parentheses around two indices denoting the symmetric and traceless part of a tensor, while brackets indicate its antisymmetric part.

\section{The Enskog and transfer equations}\lb{sec2}

In this work we shall analyse a moderately dense  gas composed of inelastic rough spherical  molecules free of external forces and torques within the framework of a kinetic theory of gases, where only binary collisions of the molecules are taken into account.

The spherical molecules are characterised by their mass $m$, diameter $\d$ and moment of inertia $I$. The linear and angular velocities of two molecules before a binary collision are denoted by $(\bc, \bc_1)$ and $(\bw, \bw_1)$, respectively, while $(\bc^\prime, \bc_1^\prime)$  and $({\bf w}^\prime$, ${\bf w}_1^\prime)$ are their corresponding velocities after collision. Let $\bk$ be a unit vector in the direction of the line which joins the centers of the molecules at collision pointing from the labeled to the unlabeled molecule and $\bg=\bc_1-\bc$  the  relative linear velocity.

For binary collisions of rough inelastic spheres a direct encounter is characterized by the pre-collisional velocities $(\bc, \bc_1,$ $ \bw, \bw_1)$,  post-collisional velocities $(\bc^\prime, \bc_1^\prime, \bw^\prime,
 \bw_1^\prime)$ and  collision vector $\bk$. For the restitution encounter the
 pre-collisional velocities, the  post-collisional velocities and the  collision vector are represented by $(\bc^\ast, \bc_1^\ast, \bw^\ast, \bw_1^\ast)$,  $(\bc, \bc_1, \bw, \bw_1)$ and $\bk^\ast=-\bk$, respectively. Furthermore,  the relationship $\bg\cdot\bk^\ast=-\alpha(\bg^\ast\cdot\bk^\ast)=-(\bg\cdot\bk)$ holds, where $\bg^\ast=\bc_1^\ast-\bc^\ast$.
The volume transformation from $(\bc^\ast, \bc_1^\ast, \bw^\ast,
 \bw_1^\ast)$,   to  $(\bc, \bc_1, \bw, \bw_1)$ is given by $
 d\bc^\ast\, d\bw^\ast\, d\bc_1^\ast\, d\bw_1^\ast=\vert J\vert
 d\bc\,d\bw\,d\bc_1\, d\bw_1,$
 where the modulus of the Jacobian of the transformation is $\vert J\vert=1/(\alpha\beta^2)$ with $\alpha$ and $\beta$ denoting the normal and tangential restitution coefficients, respectively. Hence,  the following  relationship between  the volume elements holds
 $(\bg^\ast\cdot\bk^\ast) d\bc^\ast\, d\bw^\ast\, d\bc_1^\ast\, d\bw_1^\ast=
 (\bg\cdot\bk)d\bc\,d\bw\,d\bc_1\, d\bw_1/(\alpha^2\beta^2).$

The equations  that relate the pre-collisional $(\bc, \bc_1,\bw, \bw_1)$ to the post-collisional  $(\bc', \bc_1',\bw', \bw_1')$ velocities  of the rough spherical molecules at collision read
\ben\label{1a}
 \bc^\prime&=&\bc+\widetilde\beta\left[\bg-{\d\over2}\,\bk\times(\bw+\bw_1)\right]+\left[
 \widetilde\alpha-\widetilde\beta\right](\bg\cdot\bk)\bk,
 \\\label{1b}
 \bc_1^\prime&=&\bc_1-\widetilde\beta\left[\bg-{\d\over2}\,\bk\times(\bw+\bw_1)\right]-\left[
 \widetilde\alpha-\widetilde\beta\right](\bg\cdot\bk)\bk,
 \\\label{1c}
 \bw^\prime&=&\bw-{2\widetilde\beta\over\kappa\,\d}\bk\times\left[\bg-{\d\over2}\,\bk\times(\bw+\bw_1)
 \right],
 \\\label{1d}
 \bw_1^\prime&=&\bw_1-{2\widetilde\beta\over\kappa\,\d}\bk\times\left[\bg-{\d\over2}\,
 \bk\times(\bw+\bw_1)\right].
 \een
 Furthermore, $\widetilde\alpha$ and $\widetilde\beta$ are given in terms of the normal $\alpha$ and tangential $\beta$ restitution coefficients and the dimensionless moment of inertia $\kappa=4I/m\d^2$ by 
\ben\label{2}
 \widetilde\alpha={(1+\alpha)\over2},\qquad\widetilde\beta={(1+\beta)\kappa\over2(\kappa+1)}.
 \een
 The ranges of the normal and tangential restitution coefficients are $0\leq\alpha\leq1$ and $-1\leq\beta\leq1$ and for an elastic collision of perfectly smooth spheres $\alpha=1$ and $\beta=-1$ which correspond to $\widetilde\alpha=1$ and $\widetilde\beta=0$ , whereas for an elastic encounter of perfectly rough spherical molecules $\alpha=1$ and $\beta=1$ which correspond to $\widetilde\alpha=1$ and $\widetilde\beta=\kappa/(\kappa+1)$.  The values of the dimensionless moment of inertia are in the range $0\leq\kappa\leq2/3$, where $\kappa=0$ refers to a concentration of the mass at the center of the sphere, $\kappa=2/3$ to a mass concentration at the surface of the sphere and the value $\kappa=2/5$ corresponds to a uniform distribution of the mass in the sphere.

The variation of the translational and rotational energy can be obtained from (\ref{1a}) -- (\ref{1d}) and reads
\ben\nonumber
 {m\over2}c^{\prime2}+{I\over2}w^{\prime2}+{m\over2}c_1^{\prime2}+{I\over2}w_1^{\prime2}
 -{m\over2}c^{2}-{I\over2}w^{2}-{m\over2}c_1^{2}-{I\over2}w_1^{2}
 =m\Biggl\{{\alpha^2-1\over4}(\bg\cdot\bk)^2
 \\\label{3}
  +{\beta^2-1\over4}{\kappa\over\kappa+1}\left[\bg-(\bg\cdot\bk)\bk-
  {\d\over2}\bk\times(\bw+\bw_1)\right]^2\Biggr\}.
 \een
We can infer from the above equation  that its right-hand side vanishes for elastic collisions of perfectly smooth spheres $(\alpha=1,\beta=-1)$ and for  elastic collisions of perfectly rough spherical molecules  $(\alpha=1, \beta=1)$. Hence, in those cases the total energy -- translational plus rotational energies -- is conserved at collision.

 In the phase space spanned by the space and velocity coordinates $(\bx, \bc, \bw)$  a state of a  granular gas with  rough inelastic spherical  molecules is characterized
by the one-particle distribution function $f\equiv f(\bx,\bc,\bw, t)$, such that $f (\bx,\bc,\bw, t)d\bx d\bc d\bw$ gives the number of the particles in
the volume element $d\bx$ about the position $\bx$, with linear velocities in the range $d\bc$ about $\bc$ and angular velocities in the range $d\bw$ about $\bw$. The Boltzmann equation
governs the space-time evolution of the one-particle distribution function $f$ in the phase space and the Enskog equation is  a modification of the Boltzmann equation for moderately dense gases which considers the increase in the collision probability and the volume occupied by the gas molecules (see e.g. \cite{CC,gk0}). The modification which refers to  the volume occupied by the gas molecules considers that  the product of the two distribution functions at collision should be evaluated at different points, because the molecules are separated by a distance $\pm\d\bk$, with the plus or minus sign relating  the collisions with initial velocities $(\bc^\ast,\bc_1^\ast, \bw^\ast, \bw_1^\ast)$ or $(\bc,\bc_1, \bw, \bw_1)$, respectively. The increase of the collision probability is a function  of the density of the gas at the contact point of the colliding spheres and the product of the two distribution functions are multiplied by a factor $\chi=\chi(\bx\pm\frac{\d}2\bk,t)$, with the plus and minus sign having the same meaning as before. Hence  the Enskog equation for granular gases without external forces and torques reads
 \ben\no
 {\partial f \over \partial t} +c_i{\partial f \over \partial x_i}
 =\int\bigg[{1\over\a^2\beta^2} \chi\left(\bx+\frac{\d}2\bk,t\right)f(\bx+\d\bk,\bc_1^\ast,\bw_1^\ast,t) f(\bx,\bc^\ast,\bw^\ast,t)
\\\lb{4}
  - \chi\left(\bx-\frac{\d}2\bk,t\right)f(\bx-\d\bk,\bc_1,\bw_1,t) f(\bx,\bc,\bw,t)\bigg]\,\d^2\, (\bg\cdot\bk)\,d\bk\,d{\bf c}_1d\bw_1.
 \een
thanks to the relationship between the volume elements. 

The distribution function $f$ and the factor $\chi$ are considered smooth functions of the spatial coordinate $\bx$, so that they could be approximated in Taylor series near the point $\bx$ and  by taking into account gradients up to the second order the Enskog equation  (\ref{4}) reduces to
\ben\no
 {\partial f \over \partial t} +c_i{\partial f \over\partial x_i}=\chi\int\left(\frac1{\a^2\bt^2}f_1^\ast f^\ast-f_1f\right)\d^2\, (\bg\cdot\bk)\,d\bk\,d{\bf c}_1d\bw_1+\d\int\bigg\{\chi\left(\frac1{\a^2\bt^2}\frac{\partial f_1^\ast}{\partial x_i} f^\ast+\frac{\partial f_1}{\partial x_i}f\right)
\\\no
 +\frac12\frac{\partial\chi}{\partial x_i}\left(\frac1{\a^2\bt^2}f_1^\ast f^\ast+f_1f\right)\bigg\}\,k_i\,\d^2\, (\bg\cdot\bk)\,d\bk\,d{\bf c}_1d\bw_1+ \frac{\d^2}2\int\bigg\{\chi\left(\frac1{\a^2\bt^2}\frac{\partial^2 f_1^\ast}{\partial x_ix_j} f^\ast-\frac{\partial^2 f_1}{\partial x_ix_j}f\right)
 \\\lb{5}
+\frac{\partial\chi}{\partial x_i}\left(\frac1{\a^2\bt^2}\frac{\partial f_1^\ast}{\partial x_j} f^\ast-\frac{\partial f_1}{\partial x_j}f\right)
+\frac14\frac{\partial^2 \chi}{\partial x_ix_j}\left(\frac1{\a^2\bt^2}f_1^\ast f^\ast-f_1f\right)\bigg\}\,k_i\,k_j\,\d^2\, (\bg\cdot\bk)\,d\bk\,d{\bf c}_1d\bw_1,
\een
Here $f^*_1\equiv f(\bx,\bc_1^*,\bw_1^*,t)$, $f^*\equiv f(\bx,\bc^*,\bw^*,t)$,  and so on.

The transfer equation for an arbitrary function of the velocities  is obtained from the multiplication of the Enskog equation (\ref{5}) by $\psi(\bx,\bc,\bw,t)$ and the integration of the resulting equation over all values of the velocities $\bc$ and $\bw$, yielding
\ben\lb{6}
\frac{\partial \Psi}{\partial t}+\frac{\partial}{\partial x_i}\left(\Phi_i+\Phi_i^I+\Phi_i^{II}\right)=\mathcal{P}+\mathcal{P}^I+\mathcal{P}^{II}.
\een
In the transfer equation  $\Psi$ denotes the density of an arbitrary additive quantity, $\Phi_i$ is a kinetic flux density which is associated with the flow of the molecules, $\Phi_i^{I}$ and $\Phi_i^{II}$ denote potential flux densities due to the collisional transfer of the molecules, while $\mathcal{P}$ is the kinetic production term and $\mathcal{P}^{I}$ and $\mathcal{P}^{II}$ the potential production terms. The expressions of these terms are given by
\ben\lb{7a}
&&\Psi=\int\psi fd\bc d\bw,\qquad \Phi_i=\int \psi c_i fd\bc d\bw,\qquad \Phi_i^{I}=\frac{\d}2\int\chi\left(\psi'-\psi\right)ff_1k_i\,d\Gamma,
\\\lb{7b}
&&\Phi_i^{II}=\frac{\d^2}4\int\chi\left[\left(\psi'-\psi\right)\frac{\partial}{\partial x_j}\left(\ln\frac{f}{f_1}\right)+\frac{\partial(\psi'-\psi)}{\partial x_j}\right]ff_1k_ik_j\,d\Gamma-\frac{\d^2}8\frac{\partial}{\partial x_j}\int \chi\left(\psi'-\psi\right)ff_1k_ik_j\,d\Gamma, 
\\\lb{7c}
&&\mathcal{P}=\int\left[\frac{\partial\psi}{\partial t}+c_i\frac{\partial\psi}{\partial x_i}\right]fd\bc d\bw+\int\chi\left(\psi'-\psi\right)ff_1 d\Gamma,
\\\lb{7d}
&&
\mathcal{P}^I=\frac{\d}2\int\chi\left[\left(\psi'-\psi\right)\frac{\partial}{\partial x_i}\left(\ln\frac{f}{f_1}\right)+\frac{\partial(\psi'-\psi)}{\partial x_i}\right]ff_1k_id\Gamma,
\\\no
&&\mathcal{P}^{II}=\frac{\d^2}8\int\chi\Bigg\{\frac{\partial^2(\psi'-\psi)}{\partial x_i\partial x_j}+2\frac{\partial(\psi'-\psi)}{\partial x_i}\frac{\partial}{\partial x_j}\left(\ln\frac{f}{f_1}\right)
\\\lb{7e}
&&\qquad+\left(\psi'-\psi\right)\left[\frac1{f}\frac{\partial^2f}{\partial x_i\partial x_j}+\frac1{f_1}\frac{\partial^2f_1}{\partial x_i\partial x_j}-2\frac{\partial\ln f}{\partial x_i}\frac{\partial\ln f_1}{\partial x_j}\right]
\Bigg\}ff_1k_ik_j\,d\Gamma.
\een
Here we have introduced the abbreviation $d\Gamma=\d^2\, (\bg\cdot\bk)\,d\bk\,d{\bf c}_1 d\bw_1d{\bf c} d\bw$ and for the determination of the above equations we have transformed all gradients of the factor $\chi$ into integral gradients and the latter were written as gradients of fluxes. The introduction of $\psi'$ has followed the well-known procedure by renaming   the pre-collisional velocities $(\bc^\ast, \bc_1^\ast,\bw^\ast,\bw^\ast_1)$  as $(\bc, \bc_1,\bw,\bw_1)$  and the post-collisional velocities $(\bc, \bc_1,\bw,\bw_1)$  as $(\bc', \bc_1',\bw',\bw_1')$. 

The factor $\chi$ represents the local radial distribution function whose expression in a virial expansion in terms of Pad\'e approximants reads \cite{Ree,r19}
\ben
\chi=\frac{1+0.3517\rho_*+0.0866\rho_*^2+0.0135\rho_*^3}{1-0.2733\rho_*-0.0295\rho_*^2},
\een
where $\rho_*=b\rho$ is the reduced density.

\section{A theory with 29 scalar fields}\lb{sec3}

\subsection{The balance equations}\lb{sec3a}

We are interested in analysing a kinetic theory of moderately dense granular gases with inelastic rough spheres where  a macroscopic state is characterised  by 29 scalar fields which correspond to the moments of the distribution function: mass density $\rho({\bf x},t)$, hydrodynamic velocity $v_i({\bf x},t)$, granular temperature $T({\bf x},t)$, pressure tensor $p_{ij}({\bf x},t)$, translational heat flux  $q_i({\bf x},t)$, rotational heat flux  $h_i({\bf x},t)$, hydrodynamic angular velocity $s_i({\bf x},t)$ and angular velocity flux $m_{ij}({\bf x},t)$.  In terms of the distribution function $f (\bx,\bc,\bw, t)$ these fields are defined by \cite{gk1,gk2}
\ben\lb{8a}
\rho=\int mfd\bc d\bw,\quad  v_i=\frac1\rho\int mc_ifd\bc d\bw,
\quad T={1\over3kn}\int\left[{m\over2}C^2+{I\over2} \Omega^2\right]fd\bc d\bw, \quad p_{ij}=\int m C_iC_jfd\bc d\bw,
\\\lb{8b}
q_i=\int \frac{m}2 C^2C_if d\bc d\bw, \quad h_i=\frac{I}2 \Omega^2 C_if d\bc d\bw,\quad
\quad s_i=\frac1n\int w_if d\bc d\bw,\quad m_{ij}=\int m \Omega_i C_jf d\bc d\bw,
\een
where $k$ denotes the Boltzmann constant,  $C_i=c_i-v_i$ the peculiar linear velocity and $\Omega_i=w_i-s_i$ the peculiar angular velocity.

The granular temperature $T$ is the sum of the partial translational $T_t$ and rotational $T_r$ temperatures, so that 
\ben\lb{8c}
T=\frac{T_t+T_r}2,\qquad T_t={m\over3kn}\int C^2fd\bc d\bw, \qquad T_r={I\over3kn}\int \Omega^2fd\bc d\bw.
\een

The determination of the balance equations for the 29 scalar fields are obtained from the transfer equation (\ref{6}) by specifying values for  $\psi(\bx,\bc,\bw,t)$. We are interested in  a theory with first order gradients, so that we shall leave out the terms $\Phi_i^{II}$ and $P^{II}$ in the determination of the balance equations. 

\emph{(i) Balance of mass  density ($\psi=m$)}
\ben\label{9a}
\frac{\partial \rho}{\partial t}
+\frac{\partial \rho v_i}{\partial x_i}=0.
\een
\emph{(ii) Balance of linear momentum density ($\psi=mc_i$)}
\ben\label{9b}
 \frac{\partial\rho v_i}{\partial t}
 +\frac{\partial(\rho v_iv_j+p_{ij}+p_{ij}^I)}{\partial x_j} =0,
\een
where the potential part of the pressure tensor reads
\ben\lb{9b1}
p_{ij}^I=\frac{\d}2\int \chi m(c_i'-c_i)k_jff_1d\Gamma.
\een 
\emph{(iii) Balance of total energy ($\psi=mC^2/2+I\Omega^2/2$)}
\ben\lb{9c}
&&\frac{\partial T}{\partial t}+v_j\frac{\partial T}{\partial x_j}+\frac1{3nk}\left[\frac{\partial (q_i+q_i^I)}{\partial x_i}+\frac{\partial (h_i+h_i^I)}{\partial x_i}+(p_{ij}+p_{ij}^I)\frac{\partial v_i}{\partial x_j}+\frac{I}m(m_{ij}+m_{ij}^I)\frac{\partial s_i}{\partial x_j}\right]
+T(\zeta+\zeta^I)=0,
\een
here $q_i^I$, $h_i^I$ and $m_{ij}^I$ are potential parts of the translational and rotational heat fluxes and angular velocity flux, respectively, while $\zeta$ and $\zeta^I$ refer respectively to the kinetic and potential parts of the cooling rate. Their expressions are
\ben\lb{9c1}
&&q_i^I=\frac{\d}2\int \chi \frac{m}2(C^{\prime2}-C^2)k_iff_1d\Gamma,\quad
h_i^I=\frac{\d}2\int \chi \frac{I}2(\Omega^{\prime2}-\Omega^2)k_iff_1d\Gamma,\quad m_{ij}^I=\frac{\d}2\int \chi m(w_i^{\prime}-\omega_i)k_jff_1d\Gamma,
\\\lb{9c2}
&&\zeta=-\frac1{3nkT}\int\chi\left[\frac{m}2(C^{\prime2}-C^2)+\frac{I}2(\Omega^{\prime2}-\Omega^2)
\right]f_1f\,d\Gamma
\\\lb{9c3}
&&\zeta^I=-\frac\d{6nkT}
\int \chi\left[\frac{m}2(C^{\prime2}-C^2)+\frac{I}2(\Omega^{\prime2}-\Omega^2)
\right]\frac{\partial}{\partial x_i}\left(\ln\frac{f}{f_1}\right)k_i ff_1d\Gamma.
\een
\emph{(iv) Balance of pressure tensor ($\psi=mC_iC_j$)}
\ben\lb{9d}
\frac{\partial p_{ij}}{\partial t}+\frac{\partial}{\partial x_k}\left(p_{ij}v_k+p_{ijk}+p_{ijk}^I\right)+\left(p_{ik}+p_{ik}^I\right)\frac{\partial v_j}{\partial x_k}+\left(p_{jk}+p_{jk}^I\right)\frac{\partial v_i}{\partial x_k}=P_{ij}+P_{ij}^I.
\een
Above $p_{ijk}$ and $p_{ijk}^I$ refer to the kinetic and potential parts of third-order tensors, while $P_{ij}$ and $P_{ij}^I$ the corresponding parts of production terms of the kinetic pressure tensor, respectively. They are given by
\ben\lb{9d1}
p_{ijk}=\int mC_iC_jC_kf d\bc d\bw, \qquad p_{ijk}^I=\frac\d2\int\chi m(C_i'C_j'-C_iC_j)k_kff_1 d\Gamma,
\\\lb{9d2}
P_{ij}=\int\chi m (C_i'C_j'-C_iC_j)ff_1d\Gamma,\qquad P_{ij}^I=\frac\d2\int\chi m (C_i'C_j'-C_iC_j)\frac{\partial}{\partial x_k}\left(\ln\frac{f}{f_1}\right)k_kff_1d\Gamma.
\een
\emph{(v) Balance of translational heat flux ($\psi=mC^2C_i/2$)}
\ben\lb{9e}
\frac{\partial q_i}{\partial t}+\frac{\partial (q_iv_j+q_{ij}+q_{ij}^I)}{\partial x_j}+(q_j+q_j^I)\frac{\partial v_i}{\partial x_j}+(p_{ijk}+p_{ijk}^I)\frac{\partial v_j}{\partial x_k}-\frac{p_{ij}}\rho\frac{\partial(p_{jk}+p_{jk}^I)}{\partial x_k}-\frac{p_{rr}}{2\rho}\frac{\partial(p_{ik}+p_{ik}^I)}{\partial x_k}=Q_i+Q_i^I,
\een
where $q_i^I=p_{rri}^I/2$ is the potential part of the translational heat flux, $q_{ij}$ and $q_{ij}^I$ the kinetic and potential parts of a contracted fourth-order tensor and $Q_i$ and $Q_i^I$ the kinetic and potential parts of the translational heat flux production term. Note that $q_i=p_{rri}/2$. The expressions of the new terms are
\ben\lb{9e1}
q_{ij}=\int\frac{m}{2}C^2C_iC_jf d\bc d\bw,\qquad q_{ij}^I=\frac\d2\int \chi \frac{m}{2}\left(C^{\prime 2}C_i^\prime-C^2C_i\right)k_jff_1d\Gamma,
\\\lb{9e2}
Q_i=\int\chi \frac{m}{2}\left(C^{\prime 2}C_i^\prime-C^2C_i\right)ff_1d\Gamma,\qquad Q_i^I=\frac\d2\int\chi\frac{m}{2}\left(C^{\prime 2}C_i^\prime-C^2C_i\right)\frac{\partial}{\partial x_k}\left(\ln\frac{f}{f_1}\right)k_kff_1d\Gamma.
\een
\emph{(vi) Balance of hydrodynamic angular velocity ($\psi=mw_i$)}
\ben\label{9f}
\frac{\partial \rho s_i}{\partial t}
+\frac{\partial (\rho s_iv_j+m_{ij}+m_{ij}^I)}{\partial x_j}=M_i+M_i^I.
\een
Above $M_i$ and $M_i^I$ are the kinetic and potential parts of the hydrodynamic angular velocity production terms which are defined by
\ben\lb{9f1}
 M_i=\int \chi m\left(w_i^\prime-w_i\right)ff_1d\Gamma,\qquad  M_i^I=\frac\d2\int \chi m\left(w_i^\prime-w_i\right)\frac{\partial}{\partial x_k}\left(\ln\frac{f}{f_1}\right)k_kff_1d\Gamma.
\een
\emph{(vii) Balance of angular velocity flux ($\psi=m\Omega_iC_j$)}
\ben\lb{9g}
\frac{\partial m_{ij}}{\partial t}+\frac{\partial}{\partial x_k}\left(m_{ij}v_k+m_{ijk}+m_{ijk}^I\right)+\left(m_{ik}+m_{ik}^I\right)\frac{\partial v_j}{\partial x_k}+\left(p_{jk}+p_{jk}^I\right)\frac{\partial s_i}{\partial x_k}=M_{ij}+M_{ij}^I,
\een
here $m_{ijk}$ and $m_{ijk}^I$ are the kinetic and potential parts of a third-order tensor and $M_{ij}$ and $M_{ij}^I$ the corresponding production terms. They are given by
\ben\lb{9g1}
m_{ijk}=\int m\Omega_iC_jC_kfd\bc d\bw,\qquad m_{ijk}^I=\frac\d2\int\chi m\left(\Omega_i'C_j'-\Omega_iC_j\right)k_kff_1d\Gamma,
\\\lb{9g2}
M_{ij}=\int\chi m\left(\Omega_i'C_j'-\Omega_iC_j\right)ff_1d\Gamma,\qquad M_{ij}^I=\frac\d2\int\chi m\left(\Omega_i'C_j'-\Omega_iC_j\right)\frac{\partial}{\partial x_k}\left(\ln\frac{f}{f_1}\right)k_kff_1d\Gamma.
\een
\emph{(viii) Balance of rotational heat flux ($\psi=I\Omega^2C_i/2$)}
\ben\no
\frac{\partial h_i}{\partial t}+\frac{\partial (h_iv_j+h_{ij}+h_{ij}^I)}{\partial x_j}+(h_j+h_j^I)\frac{\partial v_i}{\partial x_j}+\frac{I}m(m_{ijk}+m_{ijk}^I)\frac{\partial s_k}{\partial x_j}-\frac{I}m\frac{m_{ji}}\rho\frac{\partial(m_{jk}+m_{jk}^I)}{\partial x_k}
\\\lb{9h}
-\frac{3k}{2m}T_r\frac{\partial(p_{ik}+p_{ik}^I)}{\partial x_k}=H_i+H_i^I-\frac{I}m\frac{m_{ji}}\rho\left(M_j+M_j^I\right),
\een
where the kinetic and potential contracted fourth-order $h_{ij}$ and $h_{ij}^I$
and the corresponding production terms $H_i$ and $H_i^I$ are given by
\ben\lb{9h1}
h_{ij}=\int\frac{I}2\Omega^2C_iC_jfd\bc d\bw,\qquad h_{ij}^I=\frac\d2\int\chi\frac{I}2\left(\Omega^{\prime2}C^\prime_i-\Omega^2C_i\right)k_jff_1d\Gamma,
\\\lb{9h2}
H_i=\int\chi\frac{I}2\left(\Omega^{\prime2}C^\prime_i-\Omega^2C_i\right)ff_1d\Gamma,\qquad H_i^I=\frac\d2\int\chi\frac{I}2\left(\Omega^{\prime2}C^\prime_i-\Omega^2C_i\right)\frac{\partial}{\partial x_k}\left(\ln\frac{f}{f_1}\right)k_kff_1d\Gamma.
\een

As usual the pressure tensor is decomposed as $p_{ij}=p_{\langle ij\rangle}+p\delta_{ij}$ where $p$ denotes the hydrostatic pressure and  $p_{\langle ij\rangle}$  the shear stress which is the traceless  part of the pressure tensor. Their expressions in terms of the one-particle distribution function are
\ben\lb{10a}
p={1\over 3}
\int m C^2 f  d{\bf c}d\bw=nkT_t,\qquad p_{\langle ij\rangle}= \int m\left(
C_{ i} C_{j}-\frac13C^2\delta_{ij}\right) f  d{\bf c}d\bw,
\een
where  $T_t$  denotes the translational temperature. 

The balance equations for the partial translational $T_t$ and rotational $T_r$ temperatures follow from the balance equations for the temperature (\ref{9c}) and for the trace of the pressure tensor (\ref{9d}) and read
\ben\lb{10b}
\frac{\partial T_t}{\partial t}+v_j\frac{\partial T_t}{\partial x_j}+\frac2{3nk}\left[\frac{\partial (q_i+q_i^I)}{\partial x_i}+(p_{ij}+p_{ij}^I)\frac{\partial v_i}{\partial x_j}\right]
+T_t(\zeta_t+\zeta_t^I)=0,
\\\lb{10c}
\frac{\partial T_r}{\partial t}+v_j\frac{\partial T_r}{\partial x_j}+\frac2{3nk}\left[\frac{\partial (h_i+h_i^I)}{\partial x_i}+\frac{I}m(m_{ij}+m_{ij}^I)\frac{\partial s_i}{\partial x_j}\right]
+T_r(\zeta_r+\zeta_r^I)=0,
\een
where the partial translational $\zeta_t$, $\zeta_t^I$ and rotational $\zeta_r$, $\zeta_r^I$ cooling rates are given by
\ben\lb{10d}
\zeta_t=-\frac1{3nkT_t}\int\chi m(C^{\prime2}-C^2)f_1f\,d\Gamma,\qquad \zeta_r=-\frac1{3nkT_r}\int\chi I(\omega^{\prime2}-\omega^2)f_1f\,d\Gamma,
\\\lb{10e}
\zeta_t^I=-\frac\d{6nkT_t}
\int \chi m(C^{\prime2}-C^2)
\frac{\partial}{\partial x_i}\left(\ln\frac{f}{f_1}\right)k_i ff_1d\Gamma,\qquad
\zeta_r^I=-\frac\d{6nkT_r}
\int \chi I(\Omega^{\prime2}-\Omega^2)
\frac{\partial}{\partial x_i}\left(\ln\frac{f}{f_1}\right)k_i ff_1d\Gamma,
\een
so that the granular cooling rate is the sum of the partial granular cooling rates, namely  $T\zeta=T_t\zeta_t/2+T_r\zeta_r/2$ and $T\zeta^I=T_t\zeta_t^I/2+T_r\zeta_r^I/2$.

The  balance equations (\ref{9a}), (\ref{9b}), (\ref{9c}), (\ref{9d}), (\ref{9e}), (\ref{9f}), (\ref{9g}) and (\ref{9h}) do not represent a system of field equations for the 29 scalar fields $\rho$, $v_i$, $T$, $p_{ij}$,   $q_i$  $h_i$,  $s_i$ and  $m_{ij}$, since we have to express the constitutive quantities 
\ben\lb{11}
p_{ij}^I,\; p_{ijk},\; p_{ijk}^I,\;P_{ij},\;P_{ij}^I,\; \zeta,\;\zeta^I,\;q_i^I,\;q_{ij},\;q_{ij}^I,\;Q_i\;Q_i^I,\;m_{ij}^I,\;M_i,\;M_i^I,\;m_{ijk},\;m_{ijk}^I,\;M_{ij},\;M_{ij}^I,\;h_{ij},\;h_{ij}^I,\;H_i,\;H_i^I,
\een
which appears in these balance equations, as functions of the  29 scalar fields in order to close the system of balance equations. This will be the subject of the next section.

\subsection{The evaluation of the constitutive quantities}\lb{sec3b}

We follow \cite{r16} and introduce time-independent translational $\tau_t$  and rotational $\tau_r$ temperature rates and  the ratio of the time-independent partial temperatures 
$\theta={\tau_r}/{\tau_t}$, so that the translational and rotational temperatures are written as
\ben\lb{12a}
T_t=\tau_tT(1+\Delta),\qquad T_r=\tau_rT\left(1-\frac\Delta\theta\right).
\een
The granular temperature and the time-independent temperature rates are given by
\ben\lb{16b}
T=\frac{T_t+T_r}2,\qquad \tau_t=\frac{2}{1+\theta},\qquad \tau_r=\frac{2\theta}{1+\theta},
\een
and  $\Delta$ refers to the non-equilibrium part of the partial temperatures. In this case we write the pressure as
\ben
p=nkT_t=nk\tau_tT(1+\Delta)=p_e(1+\Delta),
\een
where $p_e$ refers to the equilibrium pressure.

From  the Maxwellian distribution function
\ben\lb{12b}
f^{(0)}=\frac{n}{(2\pi kT)^3}\left(\frac{mI}{\tau_t\tau_r}\right)^\frac32\exp\left[-{mC^2\over2k\tau_tT}-{I \Omega^2\over2k\tau_rT}\right],
\een
follows Grad's distribution function for the non-equilibrium fields $\Delta, p_{\langle ij\rangle}, q_i, h_i$ $m_{ij}$, namely
\ben\no
f=f^{(0)}\Bigg\{1+\frac{m}{2kT\tau_t}\left(C^2-\frac{I}{\theta m}\Omega^2\right)\Delta+\frac{1}{\rho}\left({m\over kT\tau_t}\right)^2\Bigg[\frac{C_iC_j}2p_{\langle ij\rangle}+C_i\left({mC^2\over 5kT\tau_t}-1\right)q_i
\\\lb{12c}
+\frac{C_i}{\theta}\left({I \Omega^2\over 3k T\tau_r}-1\right)h_i+\frac{I\Omega_iC_j }{m\theta}m_{ij}\Bigg]\Bigg\}.
\een

If we refer to the  velocity distribution function (VDF) of  Lun \cite{Lun} the  VDF (49) has new terms corresponding to the angular velocity flux $m_{ij}$ and the non-equilibrium part of the partial temperatures $\Delta$. Here the zeroth-order VDF $f^{(0)}$ is the local version of the homogeneous cooling state VDF \cite{r12,r16,r17,r20,b2}, which does not include the translation-rotation coupling  analysed in the works \cite{fr1,fr2,fr3}. The orientational correlation is an important subject which was also analysed  from MD simulation data on the VDF for shear flow of granular gases of rough spherical molecules \cite{fr3,fr4}.

From the knowledge of Grad's distribution function the constitutive equations for the kinetic and potential parts of the moments of the distribution function can be determined from the insertion of (\ref{12c}) into their definitions  (\ref{9b1}), (\ref{9c1}), (\ref{9d1}), (\ref{9e1}), (\ref{9g1}), (\ref{9h1})  and integration of the resulting equations, yielding
\ben\lb{13a}
&&p_{ij}^{I}=\chi b\rho\left[\frac{2\wa+3\wb}{5}p_{\langle ij\rangle}+\wa p_e\left(1+\Delta\right)\delta_{ij}-\d\sqrt{\frac{kT\tau_t}{\pi m}}\wb\rho\varepsilon_{ijk}s_k\right],
\\\lb{13b}
&&q_i^I=\chi b\rho\left[\frac{3\wa+2\wb}{5}q_i-\frac{\d\wb\theta}{2}\sqrt{\frac{kT\tau_t}{m\pi}}\varepsilon_{ijk}m_{jk}\right],\quad
h_i^I=\chi b \rho\left[\frac{2\wb}{3\kappa}h_i-\frac{\d\wb}2\sqrt{\frac{kT\tau_t}{m\pi}}\varepsilon_{ijk}m_{jk}\right],\quad m_{ij}^I=0,
\\\lb{13c}
&&p_{ijk}=\frac25(q_i\delta_{jk}+q_j\delta_{ik}+q_k\delta_{ij}),\qquad q_{ij}=\frac{kT\tau_t}m\left[\frac{5}2p_e\left(1+2\Delta\right)\delta_{ij}+\frac72p_{\langle ij\rangle}\right],
\\\lb{13d}
&&p_{ijk}^I=\chi b\rho\bigg\{\frac{9\wa+\wb}{25}(q_i\delta_{jk}+q_j\delta_{ik})+\frac{4\wa+6\wb}{25}q_k\delta_{ij}-\sqrt{\frac{kT\tau_t}{m\pi}}\wb\d m_{s(i}\varepsilon_{j)ks}\bigg\},
\\\no
&&q_{ij}^I=\chi b\rho\frac{p_e}\rho\Bigg\{\left[\left(\frac\wa2(8-9\wa+6\wa^2)+\wb(1-2\wa)(1-\wb)\right)(1+2\Delta)-\frac{\wb^2\theta}\kappa(1-2\wa)\right]p_e\delta_{ij}
\\\no
&&\qquad+\left[\frac\wa{10}(38-45\wa+24\wa^2)+\frac\wb5(16-19\wb+12\wb^2)+\frac{\wa\wb}5(9\wa+2\wb-11)+\frac{4\wb^2\theta}{5\kappa}(3\wb+\wa-2)\right]p_{\langle ij\rangle}
\\\lb{13e}
&&\qquad+2\d\sqrt{\frac{kT\tau_t}{\pi m}}\wb\left[\wa(1-\wa)
+2\wb\left(1-\wb-\frac{\wb\theta}\kappa\right)-\frac{11}8\right]\rho\varepsilon_{ijk}s_k\Bigg\},\qquad h_{ij}=\frac{3p_e}{2\rho}\left(p_e\delta_{ij}+p_{\langle ij\rangle}\right),
\\\no
&&m_{ijk}^I=\chi b\rho\left\{\frac\wb{5\kappa}(2\wa-1)p_e\left[s_j\delta_{ik}+s_k\delta_{ij}-4s_i\delta_{jk}\right]+\frac\wb\kappa\left(\frac{2\wb\theta}{\kappa}+6\wb-3-\theta\right)\frac{p_e}\d\sqrt{\frac{kT\tau_t}{\pi m}}\Delta\varepsilon_{ijk}\right.
\\\lb{13g}
&&\left.\qquad-\frac2{5\d}\sqrt{\frac{kT\tau_t}{\pi m}}\frac\wb\kappa\left[2\left(3-2\wa-4\wb\right)\varepsilon_{ikr}p_{\langle jr\rangle}+\left(\frac{2\wb\theta}{\kappa}+4\wa-4\wb-\theta\right)\varepsilon_{ijr}p_{\langle kr\rangle}\right]\right\},
\qquad
m_{ijk}=0,
\\\no
&& 
h_{ij}^I=\chi b\rho\frac{p_e}\rho\Bigg\{\left[\frac{\wb\theta}{\kappa}(1-2\wa)+\frac{3\wa\theta}{2}+\frac{\wb^2(2\wa-1)}{\kappa}\left(\frac\theta\kappa+1+2\Delta\right)\right]p_e\delta_{ij}
\\\no
&&\qquad+\left[\frac{3\wa\theta}5+\frac{12\wb^3}{5\kappa^2}\left(\frac\theta\kappa+1\right)+\frac{\wb^2\theta}{5\kappa}\left(2\wa\left(\frac\kappa\theta+2\right)-9\kappa-\frac{7\kappa}{\theta}-5\right)+\frac{\wb\theta}5\left(\frac92+\frac5\kappa-\frac{4\wa}\kappa\right)\right]p_{\langle ij\rangle}
\\\lb{13i}
&&\qquad-\frac{\d}2\sqrt{\frac{kT\tau_t}{\pi m}}\wb\theta\left[3+\frac{8\wb^2}{\theta\kappa}\left(1+\frac\theta\kappa\right)-\frac{2\wb}\kappa\left(\frac1\theta+3\right)\right]\rho\varepsilon_{ijk}s_k\Bigg\},
\een
In the above equations we have considered only the linear terms in the non-equilibrium fields and introduced the covolume $b=2\pi\d^3/3m$ and the effective collision frequency $\nu=4n\d^2\sqrt{\frac{\pi kT\tau_t}m}$.

The constitutive equations for the kinetic and potential production terms  (\ref{9c2}), (\ref{9c3}), (\ref{9d2}), (\ref{9e2}), (\ref{9f1}), (\ref{9g2}), (\ref{9h2}) follow the same methodology by introducing Grad's distribution function  (\ref{12c}) into their definitions, integration of the resulting equations and by considering only linear terms in the non-equilibrium fields. Hence we have
\ben\no
&&\zeta=\frac23\chi \nu\tau_t\Bigg\{\wa\left(1-\wa\right)+\wb\left(1-\wb\right)-\frac{\wb^2}{\kappa}
-\frac{\wb\theta}\kappa\left(\wb+\frac\wb\kappa-1\right)
\\\lb{14a}
&&\qquad+\frac32\left[\wa\left(1-\wa\right)+\wb\left(1-\wb\right)-\frac{\wb^2}{\kappa}
+\frac{\wb\theta}{3\kappa}\left(\wb+\frac\wb\kappa-1\right)\right]\Delta\Bigg\},
\\\no
&&\zeta^I=\chi b \rho\frac{\tau_t}{p_e}\Bigg\{\left[\wa(\wa-1)+\wb(\wb-1)+\frac{\wb^2}{\kappa}+\frac{\wb\theta}{\kappa}\left(\frac\wb\kappa+\wb-1\right)\right]p_e\frac{\partial v_k}{\partial x_k}+\frac{2\wb\theta}{9\kappa}\left(\frac\wb\kappa+\wb-1\right)\frac{\partial h_k}{\partial x_k}
\\\lb{14b}
&&\qquad+\left[\frac{\wa(\wa-1)}5+\frac{2\wb}{15}\left(\frac\wb\kappa+\wb-1\right)\right]\frac{\partial q_k}{\partial x_k}-\frac\wb3\left(\frac\wb\kappa+\wb-1\right)\d\sqrt{\frac{kT\tau_t}{\pi m}}\varepsilon_{ijk}\frac{\partial m_{ij}}{\partial x_k}\Bigg\}
\\\no
&&P_{ij}=-\nu\chi \Bigg\{p_e\left[\frac43\left(\widetilde\alpha\left(1-\widetilde\alpha\right)+\widetilde\beta\left(1-\widetilde\beta\right)-\frac{\wb^2\theta}{\kappa}\right)+2\left(\widetilde\alpha\left(1-\widetilde\alpha\right)+\widetilde\beta\left(1-\widetilde\beta\right)
+\frac{\wb^2\theta}{3\kappa}\right)\Delta\right]\delta_{ij}
\\\lb{14c}
&&\qquad+\frac{4}5\left[\frac{\widetilde\beta^2\theta}{6\kappa}-\left(\widetilde\alpha+\widetilde\beta\right)\left(\widetilde\alpha+\widetilde\beta-2\right)\right]p_{\langle ij\rangle}\Bigg\},
\\\no
&&P_{ij}^I=\chi b \rho\Bigg\{\left[\frac{6\wa}5(3-2\wa-2\wb)+\frac{2\wb}5\left(1+2\wb+\frac{2\wb\theta}\kappa\right)\right]p_e\frac{\partial v_{\langle i}}{\partial x_{j\rangle}}+\left[2\wa(1-\wa)+\frac{4\wb}{3}\left(1-\frac{\wb}{\kappa}(\theta+\kappa)\right)\right]p_e\frac{\partial v_k}{\partial x_k}\delta_{ij}
\\\no
&&+\bigg[\frac{6\wa}{25}(3-2\wa-2\wb)+\frac{2\wb}{25}(1+2\wb)\bigg]\frac{\partial q_{(i}}{\partial x_{j)}}+\bigg[\frac{2\wa}{25}(2-3\wa+2\wb)+\frac{2\wb}{25}(3-4\wb)\bigg]\frac{\partial q_k}{\partial x_k}\delta_{ij}+\frac{4\wb^2\theta}{15\kappa}\bigg[\frac{\partial h_{(i}}{\partial x_{j)}}-2\frac{\partial h_k}{\partial x_k}\delta_{ij}\bigg]
\\\lb{14d}
&&+\d\sqrt{\frac{kT\tau_t}{\pi m}}\frac\wb5\left[(4\wa+8\wb-6)\frac{\partial m_{r(i}}{\partial x_k}\varepsilon_{j)kr}+(4\wa-2\wb-1)\frac{\partial m_{rs}}{\partial x_{(i}}\varepsilon_{j)sr}\right]\Bigg\},
\\\lb{14e}
&&Q_i=\frac{\nu\chi}{15}\left\{\left[33(\wa^2+\wb^2)-41(\wa+\wb)+16\wa\wb+\frac{7\wb^2\theta}\kappa\right]q_i+\frac{50\wb^2}{3\kappa}h_i\right\},
\\\no
&&Q_i^I=\chi b \rho\frac{p_e}\rho\Bigg\{\left[\frac\wa2(11-9\wa)+\wb(2-2\wa-\wb)-\frac{\wb^2\theta}\kappa\right]\frac{p_e}T\left(\frac{\partial T}{\partial x_i}+T\frac{\partial \Delta}{\partial x_i}\right)
\\\lb{14f}
&&\qquad+\left[\frac\wa{10}(28-27\wa)+\frac\wb{10}(17-2\wa)-\frac{8\wb^2}{5}\left(1+\frac\theta\kappa\right)\right]\frac{\partial p_{\langle ij\rangle}}{\partial x_j}\Bigg\},
\\\lb{14g}
&&M_i=-\frac43\nu\chi\frac\wb\kappa\rho s_i,
\qquad
M_i^I=\chi b\rho\frac\wb\kappa\left\{\frac{2\nu\chi}{3\chi b\rho}\varepsilon_{ijk}\left[\rho\frac{\partial v_k}{\partial x_j}+\frac{m}{10k T\tau_t}\frac{\partial q_k}{\partial x_j}\right]+\frac35\frac{\partial m_{\langle ij\rangle}}{\partial x_j}+\frac{\partial m_{[ij]}}{\partial x_j}\right\},
\\\lb{14h}
&&M_{ij}=\frac23\nu\chi\left[\frac{\wb}{3\kappa}\sqrt{\frac{m\pi}{kT\tau_t}}\frac1\d\varepsilon_{ijk}h_k-\left(\frac\wb\kappa+\wa+\wb\right)m_{ij}\right],
\\\lb{14i}
&&M_{ij}^I=\chi b\rho\left\{\wa p_e\frac{\partial s_i}{\partial x_j}-\frac{2\wb}{\kappa}\frac1\d\sqrt{\frac{kT\tau_t}{\pi m}}\left[p_e\varepsilon_{ijk}\left(\frac{1+\theta}{T}\frac{\partial T}{\partial x_k}+(1-\theta)\frac{\partial \Delta}{\partial x_k}\right)+\varepsilon_{irs}\frac{\partial p_{\langle js\rangle}}{\partial x_r}\right]\right\},
\\\no
&&H_i=\nu\chi\Bigg\{\frac{\wb\theta}{3\kappa}\left(\frac\wb\kappa-1+\frac{3\wb}{\theta}\right)q_i+\frac23\left[\frac\wb\kappa\left(\frac\wb\kappa-\frac73+\frac{4\wa}{3}+2\wb\right)-(\wa+\wb)\right]h_i
\\\lb{14j}
&&\qquad+\d\sqrt{\frac{\pi kT\tau_t}{m}}\wb\left[\wb-\wa+\theta\left(\frac{\wb}{\kappa}-\frac12\right)\right]\varepsilon_{ijk}m_{jk}\Bigg\},
\\\lb{14k}
&&H_i^I=\chi b\rho\frac{p_e^2}{\rho T}\left\{\left[\frac{3\wa\theta}2+\frac{2\wb\theta}\kappa\left(1-\wa-\frac\wb{2\kappa}-\frac\wb{2\theta}\right)\right]\frac{\partial T}{\partial x_i}-\left[\frac{3\wa\theta}2+\frac{\wb\theta}\kappa\left(\frac\wb\kappa+\frac\wb\theta-2\wa\right)\right]T\frac{\partial \Delta}{\partial x_i}\right\}.
\een

For the case of elastic rough spheres $\wa=1$, $\wb=\kappa/(\kappa+1)$, $\theta=1$ and  identifying $\tau_t T\equiv T$, it follows that  the cooling rates $\zeta$ and $\zeta^I$ given by (\ref{14a}) and (\ref{14b}) vanish, the moments of the distribution function (\ref{13a}) -- (\ref{13i})   and the production terms (\ref{14c}) -- (\ref{14k}) reduce to the expressions (4.36) -- (4.57) given in \cite{gk1} (see Appendix). Here we note that in \cite{gk1} there is a misprint in the term $M_i$ related  to  a missing  factor $\kappa/(\kappa+1)$.

The  kinetic parts of the partial translational $\zeta_t$ and rotational $\zeta_r$  cooling rates  are obtained from (\ref{14a}), (\ref{14c}) and  $T\zeta=T_t\zeta_t/2+T_r\zeta_r/2$, yielding
\ben
\zeta_t=\frac43\nu\chi\left\{\widetilde\alpha\left(1-\widetilde\alpha\right)+\widetilde\beta\left(1-\widetilde\beta\right)-\frac{\wb^2\theta}{\kappa}+
\left[\widetilde\alpha\left(1-\widetilde\alpha\right)+\widetilde\beta\left(1-\widetilde\beta\right)
+\frac{3\wb^2\theta}{\kappa}\right]\frac\Delta2\right\},
\\\lb{15a}
\zeta_r=\frac43\nu\chi\frac{\wb}{\kappa\theta}\left\{\left[\theta-\wb-\frac{\wb\theta}\kappa\right]+\left[1-\frac{3}2(\wb-\theta)-\frac{\wb}\kappa\left(1+\frac{3\theta}2\right)-\wb\left(\frac1\theta+2\theta\right)\right]\Delta\right\},
\een
while the potential parts of the partial translational $\zeta_t^I$ and rotational $\zeta_r^I$ cooling rates follow from (\ref{14b}), (\ref{14d}) and  $T\zeta^I=T_t\zeta_t^I/2+T_r\zeta_r^I/2$ and read
\ben\no
\zeta_t^I=\frac{\chi b\rho}{p_e}\Bigg\{\left[2\wa(\wa-1)+\frac{4\wb}{3}\left(\frac{\wb}{\kappa}(\theta+\kappa)-1\right)\right]p_e\frac{\partial v_k}{\partial x_k}+\bigg[\frac{2\wa}{5}(\wa-1)+\frac{4\wb}{15}(\wb-1)\bigg]\frac{\partial q_k}{\partial x_k}+\frac{4\wb^2\theta}{9\kappa}\frac{\partial h_k}{\partial x_k}
\\\lb{15b}
+\d\sqrt{\frac{kT\tau_t}{\pi m}}\frac\wb{3}(1-2\wb)\varepsilon_{ijk}\frac{\partial m_{ij}}{\partial x_k}\Bigg\},
\\\no
\zeta_r^I=\frac{\chi b\rho}{p_e\theta}\Bigg\{\frac{2\wb}{3\kappa}\left[\frac\wb\kappa(3+\kappa)(\theta+\kappa)-(3\theta+\kappa)\right]p_e\frac{\partial v_k}{\partial x_k}+\frac{4\wb^2}{15\kappa}\frac{\partial q_k}{\partial x_k}+\frac{4\wb\theta}{9\kappa}\left(\frac\wb\kappa-1\right)\frac{\partial h_k}{\partial x_k}
\\\lb{15c}
-\d\sqrt{\frac{kT\tau_t}{\pi m}}\frac{2\wb}{3}\left(\frac\wb\kappa-\frac12\right)\varepsilon_{ijk}\frac{\partial m_{ij}}{\partial x_k}\Bigg\}.
\een

A nonzero stationary solution for the partial temperatures rate $\theta=\tau_r/\tau_t$ is determined from the condition $\zeta_t=\zeta_r$ when $\Delta=0$, yielding \cite{r16}
\ben\lb{16a}
\theta=\sqrt{1+h^2}+h,\qquad h=\frac{\wa\kappa^2(1-\wa)+\wb\kappa^2(1-\wb)+\wb(\wb-\kappa)}{2\wb^2\kappa}.
\een

Once the constitutive quantities (\ref{13a}) -- (\ref{14k}) are known the insertion of their expressions into the   balance equations (\ref{9a}), (\ref{9b}), (\ref{9c}), (\ref{9d}), (\ref{9e}), (\ref{9f}), (\ref{9g}) and (\ref{9h}) leads to a system of field equations for the 29 scalar fields $\rho$, $v_i$, $T$, $p_{ij}$,   $q_i$,  $h_i$,  $s_i$ and  $m_{ij}$.

 
\section{A theory with eight scalar fields}\lb{sec4}
\subsection{The balance equations}\lb{sec4a}

Let us turn our attention to a theory with eight scalar fields represented by the mass density $\rho$,  hydrodynamic velocity $v_i$,  granular temperature $T$ and hydrodynamic angular velocity $s_i$. In this case, the kinetic parts of the pressure tensor $p_{ij}$, translational heat flux  $q_i$, rotational heat flux  $h_i$ and angular velocity flux $m_{ij}$ are no longer variables and must be expressed in terms of the eight scalar fields. From now on we shall represent the kinetic parts of these fields by an index $K$. Furthermore, to have a system of field equations for the eight scalar fields the kinetic parts should be  given by constitutive equations which depend on the gradients of the eight fields, so that in the derivation of their balance equations from the transfer equation (\ref{6}) we have to take into account the potential flux density 
 $\Phi_i^{II}$ and the corresponding production term $P^{II}$. Hence, the balance equations for the eight scalar fields follow from the transfer equation (\ref{6}) by taking  the arbitrary function $\psi$ as:

\emph{(i) Balance of mass  density ($\psi=m$)}
\ben\label{17a}
\frac{\partial \rho}{\partial t}
+\frac{\partial \rho v_i}{\partial x_i}=0.
\een
\emph{(ii) Balance of linear momentum density ($\psi=mc_i$)}
\ben\label{17b}
 \frac{\partial\rho v_i}{\partial t}
 +\frac{\partial(\rho v_iv_j+p^K_{ij}+p_{ij}^I+p_{ij}^{II})}{\partial x_j} =0,
\een
where the new potential part of the pressure tensor $p_{ij}^{II}$ reads
\ben\lb{17b1}
p_{ij}^{II}=\frac{\d^2}4\int\chi m\left(c_i'-c_i\right)\frac{\partial}{\partial x_k}\left(\ln\frac{f}{f_1}\right)ff_1k_jk_k\,d\Gamma-\frac{\d^2}8\frac{\partial}{\partial x_k}\int \chi m\left(c_i'-c_i\right)ff_1k_jk_k\,d\Gamma.
\een 
\emph{(iii) Balance of temperature ($\psi=mC^2/2+I\Omega^2/2$)}
\ben\no
&&\frac{\partial T}{\partial t}+v_j\frac{\partial T}{\partial x_j}+\frac1{3nk}\bigg[\frac{\partial (q^K_i+q_i^I+q_i^{II})}{\partial x_i}+\frac{\partial (h^K_i+h_i^I+h_i^{II})}{\partial x_i}+(p^K_{ij}+p_{ij}^I+p_{ij}^{II})\frac{\partial v_i}{\partial x_j}
\\\lb{17c}
&&\qquad+I(m^K_{ij}+m_{ij}^I+m_{ij}^{II})\frac{\partial s_i}{\partial x_j}\bigg]+T(\zeta_K+\zeta^I+\zeta^{II})=0.
\een
Here the new potential parts of the translational heat flux $q_i^{II}$, rotational heat flux $h_i^{II}$, angular velocity flux $m_{ij}^{II}$ and cooling rate $\zeta^{II}$ are given by 
\ben\lb{17c1}
&&q_i^{II}=\frac{\d^2}4\int\chi \frac{m}2\left[\left(C^{\prime2}-C^2\right)\frac{\partial}{\partial x_j}\left(\ln\frac{f}{f_1}\right)+\frac{\partial(C^{\prime2}-C^2)}{\partial x_j}\right]ff_1k_ik_j\,d\Gamma-\frac{\d^2}8\frac{\partial}{\partial x_j}\int \chi \frac{m(C^{\prime2}-C^2)}2ff_1k_ik_j\,d\Gamma,\qquad
\\\lb{17c2}
&&h_i^{II}=\frac{\d^2}4\int\chi \frac{I}2\left[\left(\Omega^{\prime2}-\Omega^2\right)\frac{\partial}{\partial x_j}\left(\ln\frac{f}{f_1}\right)+\frac{\partial(\Omega^{\prime2}-\Omega^2)}{\partial x_j}\right]ff_1k_ik_j\,d\Gamma-\frac{\d^2}8\frac{\partial}{\partial x_j}\int \chi \frac{I(\Omega^{\prime2}-\Omega^2)}2ff_1k_ik_j\,d\Gamma,
\\\lb{17c3}
&&m_{ij}^{II}=\frac{\d^2}4\int\chi m\left(w_i'-w_i\right)\frac{\partial}{\partial x_k}\left(\ln\frac{f}{f_1}\right)ff_1k_jk_k\,d\Gamma-\frac{\d^2}8\frac{\partial}{\partial x_k}\int \chi m\left(w_i'-w_i\right)ff_1k_jk_k\,d\Gamma
\\\no
&&\zeta^{II}=-\frac{\d^2}{24nkT}\Bigg\{\int\chi\Bigg\{\frac{\partial^2\left[\frac{m}2(C^{\prime2}-C^2)+\frac{I}2(\Omega^{\prime2}-\Omega^2)\right]}{\partial x_i\partial x_j}+\left[\frac{m}2(C^{\prime2}-C^2)+\frac{I}2(\Omega^{\prime2}-\Omega^2)
\right]
\\\lb{17c4}
&&\qquad\times\left[\frac1{f}\frac{\partial^2f}{\partial x_i\partial x_j}+\frac1{f_1}\frac{\partial^2f_1}{\partial x_i\partial x_j}-2\frac{\partial\ln f}{\partial x_i}\frac{\partial\ln f_1}{\partial x_j}\right]\Bigg\}ff_1k_ik_jd\Gamma\Bigg\}.
\een
\emph{(iv) Balance of hydrodynamic angular velocity ($\psi=mw_i$)}
\ben\label{17d}
\frac{\partial \rho s_i}{\partial t}
+\frac{\partial (\rho s_iv_j+m^K_{ij}+m_{ij}^I++m_{ij}^{II})}{\partial x_j}=M^K_i+M_i^I+M_i^{II},
\een
where the new potential part of the  hydrodynamic angular velocity production term $M_i^{II}$ reads
\ben\lb{17d1}
M_i^{II}=\frac{\d^2}{8}\Bigg\{\int\chi m(w_i^\prime-w_i)\left[\frac1{f}\frac{\partial^2f}{\partial x_j\partial x_k}+\frac1{f_1}\frac{\partial^2f_1}{\partial x_j\partial x_k}-2\frac{\partial\ln f}{\partial x_j}\frac{\partial\ln f_1}{\partial x_k}\right]ff_1k_jk_kd\Gamma\Bigg\}.
\een

\subsection{The evaluation of the new potential constitutive  terms}\lb{sec4b}

By using the same methodology of the previous sections one can evaluate the new potential constitutive terms, yielding
\ben\lb{18a}
&&p_{ij}^{II}=-\frac{4p_e}{\nu\chi}\frac{\chi^2b^2\rho^2}\pi\left[\wa\frac{\partial v_k}{\partial x_k}\delta_{ij}+\frac{3\wb}{2}\frac{\partial v_{[i}}{\partial x_{j]}}+\frac{12\wa+9\wb}{10}\frac{\partial v_{\langle i}}{\partial x_{j\rangle}}\right],
\\\no
&&q_{i}^{II}=-\frac32\frac1{\nu\chi}\frac{p_e^2}\rho\frac{\chi^2b^2\rho^2}\pi\left[\left(\wa(1+3\wa)+\wb(1+3\wb)+\frac{3\wb^2\theta}\kappa\right)\frac{\partial\ln T}{\partial x_i}\right.
\\\lb{18b}
&&\qquad\left.-2\left(
\wa(1-\wa)+\wb(1-\wb)-\frac{\wb^2\theta}\kappa\right)\frac{\partial \ln \chi b^2\rho^2}{\partial \rho}\frac{\partial\rho}{\partial x_i}\right],
\\\lb{18c}
&&h_{i}^{II}=-\frac32\frac1{\nu\chi}\frac{p_e^2}\rho\frac{\chi^2b^2\rho^2}\pi\frac\wb\kappa\left[\left(3\wb+\theta+\frac{3\wb\theta}{\kappa}\right)\frac{\partial\ln T}{\partial x_i}
+2\left(\wb-\theta+\frac{\wb\theta}\kappa\right)\frac{\partial \ln \chi b^2\rho^2}{\partial \rho}\frac{\partial\rho}{\partial x_i}\right],
\\\lb{18d}
&&m_{ij}^{II}=\frac3{\nu\chi}p_e\frac{\chi^2b^2\rho^2}\pi\frac\wb{\kappa}\left[\frac35\frac{\partial s_{\langle i}}{\partial x_{j\rangle}}+\frac{\partial s_{[i}}{\partial x_{j]}}\right],
\\\lb{18e}
&&M_i^{II}=\frac6{5\nu\chi}p_e\frac{\chi^2b^2\rho^2}\pi\frac\wb\kappa\left(\frac{\partial^2 s_j}{\partial x_i\partial x_j}-2\frac{\partial^2 s_i}{\partial x_j\partial x_j}\right),
\\\lb{18f}
&&\zeta^{II}=\frac1{\nu\chi}\frac{p_e}\rho\frac{\chi^2b^2\rho^2}\pi\frac1{1+\theta}\left[\wa(1-\wa)+\wb(1-\wb)+\frac{\wb\theta}{\kappa}\left(1-\wb-\frac\wb\kappa-\frac\wb\theta\right)\right]\left[\frac3T\frac{\partial^2T}{\partial x_i\partial x_i}+\frac4\rho\frac{\partial^2\rho}{\partial x_i\partial x_i}\right].
\een
Note that in the above equations only linearized first order gradients were taken into account. Here we have also that for the case of elastic rough spheres where $\wa=1$, $\wb=\kappa/(\kappa+1)$, $\theta=1$ and  $\tau_t T\equiv T$,  the cooling rate $\zeta^{II}$ vanishes as well as the coefficients of the terms  in  the mass density gradient $\partial\rho/\partial x_i$ of the heat fluxes $q_i^{II}$ and $h_i^{II}$. Moreover, the terms (\ref{18a}), (\ref{18b}), (\ref{18c}) and (\ref{18d}) reduce to 
the expressions (5.11), (5.12), (5.14) e (5.15)  given in \cite{gk1}, respectively (see Appendix). In the referred work it was not considered the second order gradients in $M_i^{II}$.

\subsection{The constitutive equations for the kinetic  terms}\lb{sec4c}

For the evaluation of the constitutive equations of the kinetic terms we make use of the remaining field  equations for  the 21 scalar fields obtained from the insertion of  the constitutive equations of Section \ref{sec3b} by neglecting all non-linear terms and considering only the gradients of the basic fields $\rho, v_i, T, s_i$.

We begin by combining the field equations of the partial temperatures (\ref{10b}) and  (\ref{10c}) with the constitutive equations and obtain the following evolution equation  for  the non-equilibrium part of the partial temperatures $\Delta_K$:
\ben\lb{19a}
\frac{\partial \Delta_K}{\partial t}+\eta_1\frac{\partial v_i}{\partial x_i}=-\nu\chi \eta_2\Delta_K,
\een
where the scalar coefficients $\eta_1$ and $\eta_2$ read
\ben\lb{19b}
\eta_1=\frac{2\theta}{3(\theta+1)}\left\{1+\chi b\rho\left[\wa(3\wa-2)+\frac{\wb}{\theta\kappa^2}\left(\kappa(\kappa+3\theta)-\wb(3+\kappa)(\theta+\kappa)-2\theta\kappa(\kappa-\wb(\theta+\kappa)\right)\right]\right\},
\\\lb{19c}
\eta_2=\frac{2\theta}{3(\theta+1)}\left\{\wa(1-\wa)+\frac\wb{\theta\kappa}\left[\theta(\kappa-3)-2+\frac{\wb}{\theta\kappa}\Big(\kappa(2+3\theta^3)+\theta(2+3\kappa)+\theta^2(3+4\kappa-\kappa^2)\Big)\right]\right\}.
\een

We follow the well known methodology used in the theory of granular gases to determine the transport coefficients (see e. g. the books \cite{b1,b2,gk0}) and assume that the non-equilibrium part of the partial temperatures is proportional to the velocity divergent. According to a dimensional analysis the scalar coefficient is proportional to the inverse of the effective collision frequency, namely
\ben\lb{19d}
\Delta_K=-\widetilde\eta\frac{\partial v_i}{\partial x_i}, \qquad\hbox{where} \qquad \widetilde\eta\propto \nu\propto \frac1{\rho\sqrt{T}}.
\een 
Hence the time derivative of  the non-equilibrium part of the partial temperatures 
becomes
\ben\lb{19e}
\frac{\partial \Delta_K}{\partial t}=-\frac{\partial\widetilde\eta}{\partial T}\frac{\partial T}{\partial t}\frac{\partial v_i}{\partial x_i}-\frac{\partial\widetilde\eta}{\partial \rho}\frac{\partial \rho}{\partial t}\frac{\partial v_i}{\partial x_i}-\widetilde\eta\frac{\partial^2 v_i}{\partial t\partial x_i}=-\frac{\nu\chi\widetilde\zeta}{2}\widetilde\eta\frac{\partial v_i}{\partial x_i}+\mathcal{O}(2),
\een
 by neglecting all non-linear terms in the derivatives. Here we have used  the temperature field equation and neglected the term in the non-equilibrium part of the partial temperatures $\Delta$ from the kinetic part of the cooling rate (\ref{14a}), namely
\ben\lb{19f}
\frac{\partial T}{\partial t}=-T\nu \chi\widetilde\zeta, \qquad\hbox{where}\qquad
\widetilde\zeta=\frac23\tau_t\left[\wa\left(1-\wa\right)+\wb\left(1-\wb\right)-\frac{\wb^2}{\kappa}
-\frac{\wb\theta}\kappa\left(\wb+\frac\wb\kappa-1\right)\right].
\een
Note that in (\ref{19e}) $\mathcal{O}(2)$ refers to second order derivatives, which shall not be considered here, since we are not interested in the Burnett-type constitutive equations.

From the insertion of (\ref{19e}) into (\ref{19a}) we get the kinetic contribution to the scalar coefficient $\widetilde \eta$:
\ben\lb{19g}
\widetilde\eta=\frac{\eta_1}{\nu\chi(\eta_2+\widetilde\zeta/2)}.
\een

The evolution equation for the kinetic shear stress $p_{\langle ij\rangle}^K$ follows from the field equation of the traceless part of the pressure tensor (\ref{9d}) with the constitutive equations, yielding
\ben\lb{20a}
\frac{\partial p_{\langle ij\rangle}^K}{\partial t}+2p_e\mu_1\frac{\partial v_{\langle i}}{\partial v_{j\rangle}}=-\nu\chi\mu_2 p_{\langle ij\rangle}^K.
\een
Here the coefficients $\mu_1$ and $\mu_2$ are given by
\ben\lb{20b}
\mu_1=1+\chi b\rho\bigg[\frac{2\wa}5(3\wa+3\wb-2)-\frac\wb5\bigg(1+2\wb+\frac{2\wb\theta}\kappa\bigg)\bigg],
\qquad
\mu_2=\frac{4}5\bigg[\frac{\wb^2\theta}{6\kappa}-(\wa+\wb)(\wa+\wb-2)\bigg].
\een
Following the same methodology above we consider  the shear viscosity -- the coefficient of proportionality  of the shear stress and  the traceless part of the velocity gradient -- proportional to the square root of the absolute temperature  thanks to a dimensional analysis,  namely
\ben\lb{20c}
 p_{\langle ij\rangle}^K=-2\widetilde\mu\frac{\partial v_{\langle i}}{\partial v_{j\rangle}},\qquad \hbox{where} \qquad\widetilde\mu\propto \sqrt{T}.
\een
In this case the time derivative of the shear stress reads
\ben\lb{20d}
\frac{\partial  p^K_{\langle ij\rangle}}{\partial t}=-2\frac{\partial\widetilde\mu}{\partial T}\frac{\partial T}{\partial t}\frac{\partial v_{\langle i}}{\partial v_{j\rangle}}-2\widetilde\mu\frac{\partial^2 v_{\langle i}}{\partial v_{j\rangle}\partial t}=\nu\chi\widetilde\zeta\widetilde\mu\frac{\partial v_{\langle i}}{\partial v_{j\rangle}}+\mathcal{O}(2).
\een

From  (\ref{20a}) and (\ref{20d}) and by by neglecting the second-order gradients follows the kinetic contribution to the shear viscosity 
\ben\lb{20e}
\widetilde\mu=\frac{p_e\mu_1}{\nu\chi(\mu_2-\widetilde\zeta/2)}.
\een

The determination of the kinetic constitutive equations for the translational and rotational heat fluxes is more involved, since the production term of the rotational heat flux (\ref{14j}) depends on the angular velocity flux. Let us analyse  first the kinetic angular velocity flux evolution equation   (\ref{9g}) with the constitutive equations, which in a linearized theory with first order gradients can be written as
\ben\no
\left(1+\frac{3}{2\nu\chi}\frac{\kappa}{\wb(\kappa+1)+\wa\kappa}\frac{\partial}{\partial t}\right)m_{ij}^K=-\frac{3}{2\nu\chi}\frac{\kappa p_e}{\wb(\kappa+1)+\wa\kappa}\Bigg\{\frac13\frac{\partial s_k}{\partial x_k}\delta_{ij}+\left[1-\frac{3\wb}{5\kappa}(2\wa-1)\chi b\rho\right]\frac{\partial s_{\langle i}}{\partial x_{j\rangle}}
\\\lb{21a}
+\left[1-\frac{\wb}{\kappa}(2\wa-1)\chi b\rho\right]\frac{\partial s_{[ i}}{\partial x_{j]}}+2\frac\wb\kappa\frac1\d\sqrt{\frac{kT\tau_t}{m\pi}}(1+\theta)\chi b \rho\varepsilon_{ijk}\frac{\partial\ln T}{\partial x_k}\Bigg\}+\frac{\wb}{3[\wb(\kappa+1)+\wa\kappa]}\sqrt{\frac{m\pi}{kT\tau_t}}\frac1\d\varepsilon_{ijk}h^K_k.
\een
By using the approximation 
\ben\lb{21b}
\left(1+\frac{3}{2\nu\chi}\frac{\kappa}{\wb(\kappa+1)+\wa\kappa}\frac{\partial}{\partial t}\right)^{-1}\approx\left(1-\frac{3}{2\nu\chi}\frac{\kappa}{\wb(\kappa+1)+\wa\kappa}\frac{\partial}{\partial t}\right),
\een
and neglecting all non-linear terms in the derivatives and second order gradients we can obtain the following expression for the kinetic angular velocity flux
\ben\no
m_{ij}^K=-\frac{3}{2\nu\chi}\frac{\kappa p_e}{\wb(\kappa+1)+\wa\kappa}\Bigg\{\frac13\frac{\partial s_k}{\partial x_k}\delta_{ij}+\left[1-\frac{3\wb}{5\kappa}(2\wa-1)\chi b\rho\right]\frac{\partial s_{\langle i}}{\partial x_{j\rangle}}+\left[1-\frac{\wb}{\kappa}(2\wa-1)\chi b\rho\right]\frac{\partial s_{[ i}}{\partial x_{j]}}
\\\lb{21c}
+2\frac\wb\kappa\frac1\d\sqrt{\frac{kT\tau_t}{m\pi}}(1+\theta)\chi b \rho\varepsilon_{ijk}\frac{\partial\ln T}{\partial x_k}\Bigg\}+\underline{\frac{\wb}{3[\wb(\kappa+1)+\wa\kappa]}\sqrt{\frac{m\pi}{kT\tau_t}}\frac1\d\varepsilon_{ijk}\left(h^K_k-\frac{3}{2\nu\chi}\frac{\kappa}{\wb(\kappa+1)+\wa\kappa}\frac{\partial h_k^K}{\partial t}\right)}.
\een

From the evolution equation of the kinetic  rotational heat flux (\ref{9h}) with the constitutive equations and the expression for the kinetic angular momentum flux (\ref{21c}) we get the linearized equation
\ben\lb{22a}
\varphi\frac{\partial h_i^K}{\partial t}+\frac{p_e^2}{\rho T}\lambda_1^R\frac{\partial T}{\partial x_i}
+\frac{p_e^2}{\rho^2}\vartheta_R\frac{\partial \rho}{\partial x_i}+\frac{p_e^2}{\rho\nu\chi}\xi_R\varepsilon_{ijk}\frac{\partial s_j}{\partial x_k}
=-\nu\chi\left(\lambda_3^T q^K_i+\lambda_2^Rh^K_i\right),
\een
where the scalar coefficients read
\ben\lb{22b}
&&\varphi=1+\frac{\pi\wb^2\kappa}{6[\wb(\kappa+1)+\wa\kappa]^2}\left[\wb-\wa+\theta\left(\frac\wb\kappa-\frac12\right)\right],
\\\lb{22c}
&&\lambda_1^R=\frac32\bigg\{1+\frac23\chi b \rho\bigg[\frac{3\wa}2(\theta-1)-\frac{2\wa\wb\theta}{\kappa}+\frac{\wb^2}\kappa\left(\frac\theta\kappa+1\right)(4\wa-1)+\frac{\wb^2(1+\theta)}{\wb(\kappa+1)+\wa\kappa}\bigg(\wb-\wa+\theta\bigg(\frac\wb\kappa-\frac12\bigg)\bigg)\bigg]\bigg\},\qquad
\\\lb{22d}
&&\vartheta_R=\chi b\rho \left(1+\frac{\partial\ln\chi b\rho}{\partial \ln\rho}\right)\bigg[\frac{3\wa}2(\theta-1)-(2\wa-1)\frac{\wb}\kappa\bigg(\frac{\wb\theta}{\kappa}+\wb-\theta\bigg)\bigg],\qquad
\lambda_3^T=-\frac{\wb\theta}{3\kappa}\left(\frac\wb\kappa-1+\frac{3\wb}{\theta}\right),
\\\no
&&\xi_R=\chi b\rho\bigg\{\frac{3\wb\kappa}{2(\wb(\kappa+1)+\wa\kappa)}\left[\wb-\wa+\theta\left(\frac\wb\kappa-\frac12\right)\right]\left[1-\frac\wb\kappa(2\wa-1)\chi b \rho\right]
\\\lb{22e}
&&\qquad+\frac{\chi b\rho}\pi3\wb\bigg[3(\theta-1)+\frac{8\wb^2}\kappa\bigg(1+\frac\theta\kappa\bigg)-\frac{2\wb}\kappa(3\theta+1)\bigg]\bigg\},
\\\lb{22f}
&&\lambda_2^R=-\frac23\left[\frac\wb\kappa\left(\frac\wb\kappa-\frac73+\frac{4\wa}3+2\wb\right)-\wa-\wb\right]-\frac{\pi\wb^2}{9[\wb(\kappa+1)+\wa\kappa]}\left[\wb-\wa+\theta\left(\frac\wb\kappa-\frac12\right)\right].
\een

The evolution equation of the kinetic translational heat flux follows from (\ref{9e}) with the constitutive equations, yielding
\ben\lb{23a}
\frac{\partial q_i^K}{\partial t}+\frac{p_e^2}{\rho T}\lambda_1^T\frac{\partial  T}{\partial x_i}
+\frac{p_e^2}{\rho^2}\vartheta_T\frac{\partial \rho}{\partial x_i}
+\frac{p_e^2}{\rho\nu\chi}\xi_T\varepsilon_{ijk}\frac{\partial s_j}{\partial x_k}
=-\nu\chi\left(\lambda_2^T q^K_i+\lambda_3^Rh^K_i\right).
\een
Here the scalar coefficients are given by
\ben\lb{23b}
&&\lambda_1^T=\frac52\left\{1+\frac{2\chi b \rho}{5}\left[\frac{3\wa^2}2(4\wa-3)-2\wa\wb+\wb^2(4\wa-1)\left(\frac\theta\kappa+1\right)\right]\right\},
\\\lb{23c}
&&\vartheta_T= \chi b\rho\left(1+\frac{\partial\ln\chi b\rho}{\partial \ln\rho}\right)\left[\frac\wa2(3-9\wa+6\wa^2)+(1-2\wa)\wb\left(1-\wb-\frac{\wb\theta}\kappa\right) \right],
\\\lb{23d}
&&\xi_T=\frac{\chi^2 b^2\rho^2}\pi12\wb\left[\frac{21}8+\wa(\wa-1)+2\wb\left(\frac{\wb\theta}\kappa+\wb-1\right)\right],
\\\lb{23e}
&&\lambda_2^T=-\frac1{15}\left[33(\wa^2+\wb^2)-41(\wa+\wb)+16\wa\wb+\frac{7\wb^2\theta}{\kappa}\right],
\qquad
\lambda_3^R=-\frac{10}9\frac{\wb^2}\kappa.
\een

We can infer from the kinetic evolution equations of the rotational (\ref{22a}) and translational (\ref{23a}) heat fluxes that they depend on the thermodynamic forces: temperature gradient, mass density gradient and the curl of the spin velocity. Hence we consider that the kinetic rotational  and translational  heat fluxes have a linear dependence on these thermodynamic forces, namely
\ben\lb{24a}
h_i^K=-\widetilde\lambda_R\frac{\partial  T}{\partial x_i}-\widetilde\vartheta_R\frac{\partial  \rho}{\partial x_i}-\widetilde \xi_R\varepsilon_{ijk}\frac{\partial s_j}{\partial x_k},\qquad q_i^K=-\widetilde\lambda_T\frac{\partial  T}{\partial x_i}-\widetilde\vartheta_T\frac{\partial  \rho}{\partial x_i}-\widetilde \xi_T\varepsilon_{ijk}\frac{\partial s_j}{\partial x_k}.
\een
Furthermore, the dependence of the scalar coefficients on the absolute temperature is assumed by dimensional analysis as: 
\ben\lb{24b}
\widetilde\lambda_R\propto \sqrt{T},\qquad\widetilde\lambda_T\propto \sqrt{T},\qquad\widetilde\vartheta_R\propto \frac{\sqrt{T^3}}\rho,\qquad\widetilde\vartheta_T\propto \frac{\sqrt{T^3}}\rho,\qquad\widetilde\xi_R\propto \frac{T}\rho,\qquad\widetilde\xi_T\propto \frac{T}\rho,
\een
so that the time derivative of the kinetic rotational and translational heat fluxes become
\ben\no
\frac{\partial h_i^K}{\partial t}=-\frac{\partial \widetilde \lambda_R}{\partial T}\frac{\partial T}{\partial t}\frac{\partial  T}{\partial x_i}-\frac{\partial \widetilde \vartheta_R}{\partial T}\frac{\partial T}{\partial t}\frac{\partial  \rho}{\partial x_i}-\frac{\partial \widetilde \vartheta_R}{\partial\rho}\frac{\partial \rho}{\partial t}\frac{\partial  \rho}{\partial x_i}-\frac{\partial \widetilde \xi_R}{\partial T}\frac{\partial T}{\partial t}\varepsilon_{ijk}\frac{\partial s_j}{\partial x_k}-\frac{\partial \widetilde \xi_R}{\partial \rho}\frac{\partial \rho}{\partial t}\varepsilon_{ijk}\frac{\partial s_j}{\partial x_k}-\widetilde\lambda_R\frac{\partial^2  T}{\partial t\partial x_i}
\\\lb{24c}
-\widetilde\vartheta_R\frac{\partial^2  \rho}{\partial t\partial x_i}-\widetilde \xi_R\varepsilon_{ijk}\frac{\partial^2 s_j}{\partial t\partial x_k}=2\nu\chi\widetilde\zeta\widetilde \lambda_R\frac{\partial  T}{\partial x_i}+\left[\frac32\widetilde\vartheta_R+\frac{\widetilde\lambda_RT}{\rho}\left(1+\frac{\partial\ln\chi}{\partial\ln\rho}\right)\right]\nu\chi\widetilde\zeta\frac{\partial  \rho}{\partial x_i}+\nu\chi\widetilde\zeta\widetilde \xi_R\varepsilon_{ijk}\frac{\partial s_j}{\partial x_k}+\mathcal{O}(2),
\\\lb{24d}
\frac{\partial q_i^K}{\partial t}
=2\nu\chi\widetilde\zeta\widetilde \lambda_T\frac{\partial  T}{\partial x_i}+\left[\frac32\widetilde\vartheta_T+\frac{\widetilde\lambda_TT}{\rho}\left(1+\frac{\partial\ln\chi}{\partial\ln\rho}\right)\right]\nu\chi\widetilde\zeta\frac{\partial  \rho}{\partial x_i}+\nu\chi\widetilde\zeta\widetilde \xi_T\varepsilon_{ijk}\frac{\partial s_j}{\partial x_k}+\mathcal{O}(2).
\een
Here we have eliminate the time derivative of the absolute temperature from (\ref{19f}) and neglected all non-linear terms in the derivatives and second order derivatives.

Now one can obtain the kinetic contributions  to the coefficients of the thermodynamic forces of the kinetic translational and rotational heat fluxes which follow from (\ref{22a}), (\ref{23a}), (\ref{24c}) and (\ref{24d}), yielding
\ben\lb{25a}
\widetilde\lambda_T=\frac{p_e^2}{\rho T\nu\chi}\frac{\lambda_1^R\lambda_3^R-\lambda_1^T(\lambda_2^R-2\widetilde\zeta\varphi)}{\lambda_3^T\lambda_3^R-(\lambda_2^T-2\widetilde\zeta)(\lambda_2^R-2\widetilde\zeta\varphi)},\qquad \widetilde\lambda_R=\frac{p_e^2}{\rho T\nu\chi}\frac{\lambda_1^T\lambda_3^T-\lambda_1^R(\lambda_2^T-2\widetilde\zeta)}{\lambda_3^T\lambda_3^R-(\lambda_2^T-2\widetilde\zeta)(\lambda_2^R-2\widetilde\zeta\varphi)},
\\\lb{25b}
\widetilde\vartheta_T=\frac{2p_e^2}{\rho^2\nu\chi}\left\{\frac{2\vartheta_R\lambda_3^R-\vartheta_T(2\lambda_2^R-3\widetilde\zeta\varphi)}{4\lambda_3^R\lambda_3^T-(2\lambda_2^T-3\widetilde\zeta)(2\lambda_2^R-3\widetilde\zeta\varphi)}+\frac{T\rho}{p_e^2}\left(1+\frac{\partial\ln\chi}{\partial\ln\rho}\right)\nu\chi\widetilde\zeta\frac{2\lambda_3^R\widetilde\lambda_R\varphi-\widetilde\lambda_T(2\lambda_2^R-3\widetilde\zeta\varphi)}{4\lambda_3^R\lambda_3^T-(2\lambda_2^T-3\widetilde\zeta)(2\lambda_2^R-3\widetilde\zeta\varphi)}\right\},
\\\lb{25c}
\widetilde\vartheta_R=\frac{2p_e^2}{\rho^2\nu\chi}\left\{\frac{2\vartheta_T\lambda_3^T-\vartheta_R(2\lambda_2^T-3\widetilde\zeta)}{4\lambda_3^R\lambda_3^T-(2\lambda_2^T-3\widetilde\zeta)(2\lambda_2^R-3\widetilde\zeta\varphi)}+\frac{T\rho}{p_e^2}\left(1+\frac{\partial\ln\chi}{\partial\ln\rho}\right)\nu\chi\widetilde\zeta\frac{2\lambda_3^T\widetilde\lambda_T-\widetilde\lambda_R\varphi(2\lambda_2^T-3\widetilde\zeta)}{4\lambda_3^R\lambda_3^T-(2\lambda_2^T-3\widetilde\zeta)(2\lambda_2^R-3\widetilde\zeta\varphi)}\right\},
\\\lb{25d}
\widetilde\xi_T=\frac{p_e^2}{\rho \nu^2\chi^2}\frac{\xi_R\lambda_3^R-\xi_T(\lambda_2^R-\widetilde\zeta\varphi)}{\lambda_3^T\lambda_3^R-(\lambda_2^T-\widetilde\zeta)(\lambda_2^R-\widetilde\zeta\varphi)},\qquad \widetilde\xi_R=\frac{p_e^2}{\rho T\nu^2\chi^2}\frac{\xi_T\lambda_3^T-\xi_R(\lambda_2^T-\widetilde\zeta)}{\lambda_3^T\lambda_3^R-(\lambda_2^T-\widetilde\zeta)(\lambda_2^R-\widetilde\zeta\varphi)}.
\een

In the limit of elastic rough spheres -- where $\wa=1$, $\wb=\kappa/(\kappa+1)$ -- the kinetic contributions of the coefficients $\widetilde\eta$, $\widetilde \mu$,   $\widetilde\lambda_T$ and $\widetilde\lambda_R$ reduce to the expressions given in (5.21), (5.22), (5.23) and (5.25) of reference \cite{gk1}, but there is a difference in the  kinetic contributions of the coefficients $\widetilde\xi_T$ and $\widetilde\xi_R$,  which is a consequence that some terms in equation (5.20) related with the  curl of the spin velocity  $\varepsilon_{ijk}\frac{\partial s_j}{\partial x_k}$ are missing in this reference. Furthermore,  the coefficients $\widetilde\vartheta_T$ and $\widetilde\vartheta_R$ vanish in the limit of elastic rough spheres, since there is no contribution of the mass density gradient  to the heat fluxes in the case for  elastic
encounters of perfectly rough spherical molecules.

\subsection{The Navier-Stokes-Fourier constitutive equations}

From the knowledge of the kinetic and potential contributions to  the constitutive equations  the system of balance equations (\ref{17a}), (\ref{17b}),  (\ref{17c}), and (\ref{17d}) becomes a system of field equations for the basic fields $\rho$, $v_i$, $T$ and $s_i$
which we reproduce here
\ben\lb{26a}
\frac{\partial \rho}{\partial t}
+\frac{\partial  \rho v_i}{\partial x_i}=0,
\qquad
 \frac{\partial\rho v_i}{\partial t}
 +\frac{\partial(\rho v_iv_j+p_{ij}^{KP})}{\partial x_j} =0,
\\\lb{26b}
\frac{\partial T}{\partial t}+v_j\frac{\partial T}{\partial x_j}+\frac1{3nk}\bigg[\frac{\partial q^{KP}_i}{\partial x_i}+p^{KP}_{ij}\frac{\partial v_i}{\partial x_j}
+Im^{KP}_{ij}\frac{\partial s_i}{\partial x_j}\bigg]+T\zeta^{KP}=0,
\\
\frac{\partial \rho s_i}{\partial t}
+\frac{\partial (\rho s_iv_j+m^{KP}_{ij})}{\partial x_j}=M^{KP}_i,
\een
where the subscript $KP$ refers to the sum of the kinetic and potential contributions, namely, $p_{ij}^{KP}=p_{ij}^{K}+p_{ij}^{I}+p_{ij}^{II}$, $q_i^{KP}=q_i^{K}+q_i^{I}+q_i^{II}+h_i^{K}+h_i^{I}+h_i^{II}$ and $m_{ij}^{KP}=m_{ij}^{K}+m_{ij}^{I}+m_{ij}^{II}$. Note that the total heat flux is the sum of the translational and rotational heat fluxes.

From the sum of the contributions given by equations (\ref{13a}), (\ref{18a}), (\ref{19d}), (\ref{19g}), (\ref{20c}) and (\ref{20e}) we get that the total pressure tensor reads
\ben\lb{pp1}
&&p_{ij}^{KP}=p_e(1+\wa\chi b\rho)\delta_{ij}-\eta\frac{\partial v_k}{\partial x_k}\delta_{ij}
-2\mu\frac{\partial v_{\langle i}}{\partial x_{j\rangle}}+\varpi\varepsilon_{ijk}\left[\frac12({\rm curl}\,{\bf v})_k-s_k\right],
\een
where the coefficients of bulk viscosity $\eta$, shear viscosity $\mu$ and rotational viscosity $\varpi$ are given by
\ben
\eta=\left[p_e(1+\wa\chi b\rho)\widetilde\eta+\frac{4p_e}{\nu\chi}\frac{\chi^2b^2\rho^2}{\pi}\wa\right],\qquad \varpi=\frac{6p_e}{\nu\chi}\frac{\chi^2b^2\rho^2}{\pi}\wb,
\\
\mu=\left[\widetilde\mu\left(1+\chi b\rho\frac{2\wa+3\wb}{5}\right)+\frac{p_e}{\nu\chi}\frac{\chi^2b^2\rho^2}{\pi}\frac{12\wa+9\wb}{5}\right].
\een 
The total equilibrium pressure is given by $p_e(1+\wa\chi b\rho)$.

By collecting all contributions to the total heat flux (\ref{13b}), (\ref{18b}), (\ref{18c}), (\ref{24a}), (\ref{25a}) -- (\ref{25d}) we get 
\ben\lb{pp2}
q_{i}^{KP}=-\lambda\frac{\partial T}{\partial x_i}-\vartheta\frac{\partial \rho}{\partial x_i}-\xi\varepsilon_{ijk}\frac{\partial s_j}{\partial x_k},
\een
where the coefficient of thermal conductivity $\lambda$, the Dufour-like coefficient $\vartheta$  and the coefficient of thermal-mechanical coupling $\xi$ read 
\ben\no
&&\lambda=\Bigg[1+\chi b\rho \frac{3\wa+2\wb}{5}\Bigg]\widetilde\lambda_T+\Bigg[1+\chi b\rho \Bigg(\frac{2\wb}{3\kappa}-\frac{\wb^2(\theta+1)}{3[\wb(\kappa+1)+\wa\kappa]}\Bigg)\Bigg]\widetilde\lambda_R
\\
&&\qquad+\frac{3p_e^2}{2\nu\chi\rho T}\frac{\chi^2b^2\rho^2}{\pi}\Bigg[\wa(1+3\wa)+\wb(1+3\wb)+\frac\wb\kappa\left(3\wb(1+\theta)+\theta +\frac{3\wb\theta}\kappa\right)-\frac{2\wb^2(\theta+1)^2}{\wb(\kappa+1)+\wa\kappa}\Bigg],
\\\no
&&\vartheta=\left(1+\chi b\rho \frac{3\wa+2\wb}{5}\right)\widetilde\vartheta_T+\left[1+\chi b\rho \left(\frac{2\wb}{3\kappa}-\frac{\wb^2(\theta+1)}{3[\wb(\kappa+1)+\wa\kappa]}\right)\right]\widetilde\vartheta_R
\\
&&\qquad+\frac{3p_e^2}{\nu\chi\rho^2 }\frac{\chi^2b^2\rho^2}{\pi}\Bigg[\wa(1-\wa)+\wb(1-\wb)-\frac\wb\kappa\Bigg(\wb(\theta+\wb)-\theta+\frac{\wb\theta}\kappa\Bigg)\Bigg]\frac{\partial\ln\chi b^2\rho^2}{\partial\ln\rho},
\\\no
&&\xi=\chi b\rho \left(1+\frac{3\wa+2\wb}{5}\right)\widetilde\xi_T+\left[1+\chi b\rho \left(\frac{2\wb}{3\kappa}-\frac{\wb^2(\theta+1)}{3[\wb(\kappa+1)+\wa\kappa]}\right)\right]\widetilde\xi_R
\\
&&\qquad-\frac{p_e^2}{\rho\nu^2\chi^2}\frac{\chi^2b^2\rho^2}{\pi}\frac{9\wb(\theta+1)\kappa}{2[\wb(\kappa+1)+\wa\kappa]}\left[1-\frac\wb\kappa(2\wa-1)\chi b\rho\right].
\een

The total angular velocity flux follows from (\ref{13b})$_3$, (\ref{18d}), (\ref{21c}), (\ref{24a})$_1$ and reads
\ben\lb{pp3}
m_{ij}^{KP}=-\sigma_1\frac{\partial s_{\langle i}}{\partial x_{j\rangle}}-\sigma_2\frac{\partial s_{k}}{\partial x_{k}}\delta_{ij}-\sigma_3\frac{\partial s_{[ i}}{\partial x_{j]}}-\sigma_4\varepsilon_{ijk}\frac{\partial T}{\partial x_{k}}-\sigma_5\varepsilon_{ijk}\frac{\partial \rho}{\partial x_{k}},
\een
where $\sigma_i (i=1,2,3)$ are associated with coefficients of  angular velocity gradients, $\sigma_4$ to a mechanical-thermal coupling and $\sigma_5$ to a density gradient. Their expressions are  
\ben
\sigma_1=\frac{3p_e}{2\nu\chi}\left\{\frac{\kappa}{\wb(\kappa+1)+\wa\kappa}\left[1-\frac{3\wb}{5\kappa}(2\wa-1)\chi b\rho\right]-\frac{6\wb}{5\kappa}\frac{\chi^2b^2\rho^2}{\pi}\right\},\qquad \sigma_2=\frac{p_e}{2\nu\chi}\frac{\kappa}{\wb(\kappa+1)+\wa\kappa},
\\
\sigma_3=\frac{3p_e}{2\nu\chi}\left\{\frac{\kappa}{\wb(\kappa+1)+\wa\kappa}\left[1-\frac{\wb}{\kappa}(2\wa-1)\chi b\rho\right]-\frac{2\wb}{\kappa}\frac{\chi^2b^2\rho^2}{\pi}+\frac{\wb\pi}{27[\wb(\kappa+1)+\wa\kappa]}\frac{\nu^2\chi^2}{p_e^2\chi b\rho}\widetilde\xi_R\right\},
\\
\sigma_4=\frac{\wb}{\wb(\kappa+1)+\wa\kappa}\left[\frac{\nu\chi\pi}{18p_e\chi b}\widetilde\lambda_R+\frac{p_e}T\frac{1+\theta}2\right],\qquad \sigma_5=\frac{\wb}{\wb(\kappa+1)+\wa\kappa}\frac{\nu\chi\pi}{18p_e\chi b}\widetilde\vartheta_R.
\een

For the cooling rate and the angular velocity production term  we consider only the 
terms which lead to first-order gradients. Hence from   (\ref{14a}), (\ref{14b}) and  (\ref{14h}) we get  for $\zeta^{KP}=\zeta^{K}+\zeta^{I}$ and $M_i^{KP}=M_i^{K}+M_i^{I}$ the following expressions
\ben
&&\zeta^{KP}=\chi \nu\tau_t\left[\wa\left(1-\wa\right)+\wb\left(1-\wb\right)-\frac{\wb^2}{\kappa}
-\frac{\wb\theta}\kappa\left(\wb+\frac\wb\kappa-1\right)\right]\left[\frac23-\left(\widetilde\eta+\frac1{\nu\chi}\right)\chi b \rho\frac{\partial v_i}{\partial x_i}\right],
\\
&&M_i^{KP}=\frac{4\wb}{3\kappa}\nu\chi\rho\left[\frac12({\rm curl}\,{\bf v})_i-s_i\right].
\een

\subsection{The limiting case of elastic rough spheres }

In the limit of elastic rough spheres we have $\wa=1$, $\wb=\kappa/(\kappa+1)$ and the transport coefficients reduce to

\emph{(i) Bulk viscosity }
\ben
\eta=\frac{4p_e}{\nu\chi}\left[\frac{(1+\kappa)^2}{32\kappa}(1+\chi b\rho)^2+\frac{\chi^2b^2\rho^2}\pi\right].
\een

\emph{(ii) Shear viscosity }
\ben
\mu=\frac{3p_e}{\nu\chi}\left[\frac{5(1+\kappa)^2}{2(13\kappa+6)}
\left(1+\frac{5\kappa+2}{5(\kappa+1)}\chi b\rho\right)^2+\frac{7\kappa+4}{5(\kappa+1)}\frac{\chi^2b^2\rho^2}\pi\right].
\een

\emph{(iii) Rotational viscosity }
\ben
 \varpi=\frac{3p_e}{\nu\chi}\frac{\chi^2b^2\rho^2}\pi\frac\kappa{\kappa+1}.
\een

\emph{(iv) Thermal conductivity }

\ben
&&\lambda=\frac{3p_e^2}{4\rho T\nu\chi}\left[\frac{\Lambda_1}{\Lambda_2} +\frac{8(2\kappa^2+3\kappa+2)}{(\kappa+1)^2}\frac{\chi^2\b^2\rho^2}{\pi}\right],
\\\no
&&\Lambda_1=(1+\kappa)^4\left[25\pi\kappa+24(37+151\kappa+50\kappa^2)\right]+2(1+\kappa)^2\big[5\pi\kappa(3+8\kappa+5\kappa^2)+8(69+419\kappa
\\\no
&&\qquad+601\kappa^2+465\kappa^2+150\kappa^3)\big]\chi b\rho+\big[\pi\kappa(9+48\kappa+94\kappa^2+80\kappa^3+25\kappa^4)+8(43+342\kappa
\\
&&\qquad+886\kappa^2+1289\kappa^3+1169\kappa^4+630\kappa^5+150\kappa^6)\big]\chi^2b^2\rho^2,
\\
&&\Lambda_2=(1+\kappa)^2\big[\pi\kappa(4+17\kappa)+8(12+75\kappa+101\kappa^2+102\kappa^3)\big].
\een

\emph{(v) Dufour-like coefficient} $\vartheta=0$.

\emph{(vi) Coefficient of thermal-mechanical coupling }
\ben
&&\xi=\frac{9p_e^2}{4\nu^2\chi^2\rho}\left[\frac{X_1}{X_2}+\frac{2(1+\kappa-\chi b\rho)}{(1+\kappa)^2}\frac{\chi^2\b^2\rho^2}{\pi}\right],
\\\no
&&X_1=\kappa\chi b\rho\big\{-8\pi(1+\kappa)^2(3+19\kappa)+(1+\kappa)[\pi(8+175\kappa+248\kappa^2+55\kappa^3)+2520(1+\kappa)^3(1+2\kappa)]\chi b\rho
\\
&&\qquad+[\pi(16+169\kappa+345\kappa^2+323\kappa^3+105\kappa^4)+168\kappa(9+52\kappa+86\kappa^2+78\kappa^3+34\kappa^4)]\chi^2\b^2\rho^2\big\},
\\
&&X_2=(1+\kappa)^2\big[\pi\kappa(4+17\kappa)+8(12+75\kappa+101\kappa^2+102\kappa^3)\big].
\een

\emph{(vi) Coefficients of  angular velocity gradients}
\ben
&&\sigma_1=\frac{3p_e}{4\nu\chi}\left[1-\frac3{5(\kappa+1)}\chi b\rho -\frac{12}{5(\kappa+1)}\frac{\chi^2b^2\rho^2}{\pi}\right],\qquad \sigma_2=\frac{p_e}{4\nu\chi},
\\
&&\sigma_3=\frac{3p_e}{4\nu\chi}\bigg\{1-\frac{\chi b\rho}{\kappa+1}\left(1+\frac4\pi\chi b\rho\right)-\frac{\kappa[\pi(4+21\kappa+17\kappa^2)-(4\pi+420\kappa+17\pi\kappa+420\kappa^2)\chi b\rho]}{(1+\kappa)\big[\pi\kappa(4+17\kappa)+8(12+75\kappa+101\kappa^2+102\kappa^3)\big]}\bigg\}.
\een

\emph{(vii) Coefficient of  mechanical-thermal coupling}
\ben
\sigma_4=\frac{p_e}{2T\chi b\rho}\frac{4\pi(1+\kappa)^2(19\kappa+3)+(96+8\pi+600\kappa+57\pi\kappa+808\kappa^2+74\pi\kappa^2+816\kappa^3+25\pi\kappa^3)\chi b\rho}{(1+\kappa)\big[\pi\kappa(4+17\kappa)+8(12+75\kappa+101\kappa^2+102\kappa^3)\big]}.
\een

\emph{(viii) Coefficient associated with the density gradient} $\sigma_5=0$.

The coefficients of bulk $\eta$, shear $\mu$ and rotational $\varpi$ viscosities as well the coefficients of angular velocity gradients $\sigma_1$ and $\sigma_2$  matches with the results of \cite{gk1}. There are small differences in the factors associated with $\chi b\rho$ for the  thermal conductivity $\lambda$, thermal-mechanical coupling $\xi$, mechanical-thermal coupling $\sigma_4$ and the coefficients of angular velocity gradients $\sigma_3$ and $\sigma_4$, since in \cite{gk1} it was not considered the underlined term in (\ref{21c}) for the determination of the transport coefficients. Note that in the elastic case the coefficient associated with the density gradient vanish.

As was pointed out in \cite{gk1} the constitutive equations for the total pressure tensor (\ref{pp1}), total heat flux (\ref{pp2}) and total angular velocity flux (\ref{pp3}) have the same dependence on the gradients of hydrodynamic velocity, angular velocity and temperature as those of the phenomenological theory of a polar fluid where the constitutive equations for the pressure tensor and angular velocity flux are non-symmetric tensors \cite{Cow,Les}.

The terminology adopted above is that the mechanical-thermal coupling coefficient is the  propotionality coefficient between the total angular velocity flux and the temperature gradient, while the thermal-mechanical coupling coefficient is the propotionality coefficient between the total heat flux and the angular velocity gradient.

\section{Conclusions}
In this work a moderately dense gas of inelastic rough spherical molecules, free of external forces and torques, was analysed within a kinetic theory based on the Boltzmann-Enskog equation where only binary collisions of the molecules were taken into account. 

A macroscopic state of the gas was characterised by 29 scalar fields which are the moments of the distribution function: mass density, hydrodynamic velocity, granular temperature, pressure deviator, translational and rotational heat fluxes, hydrodynamic angular velocity and angular velocity flux. The balance equations for the basic fields were determined from a transfer equation derived from the Boltzmann-Enskog equation. The constitutive terms of the balance equations were obtained from Grad's distribution function expressed with respect to the 29 scalar fields.

From the knowledge of the 29 scalar field equations a transition to a eight scalar fields (mass density, hydrodynamic velocity, granular temperature and hydrodynamic angular velocity) was done and the new kinetic constitutive terms were calculated from the remaining field equations. The Navier-Stokes-Fourier constitutive equations and the transport coefficients  were identified and the limiting case of the transport coefficients corresponding to the conservative elastic rough spheres was obtained. 

With respect to the transport coefficients there are several cases to be analysed, the cases of pure smooth when $\wb=0$, quasi-smooth when $\wb\rightarrow0$,  the behaviour of the transport coefficients for different values of the normal $\alpha$ and tangential $\beta$ restitution coefficients and their dependence on the local equilibrium radial distribution function $\chi$, which is related with the increase of the gas density. Another subject that it is important to analyse is the instability of the homogeneous cooling state. These topics  will be the subject of a forthcoming paper.

\section*{Appendix}

In this appendix we give the expressions for the moments of the distribution function and production terms for the case of elastic rough spheres where $\wa=1$, $\wb=\kappa/(\kappa+1)$, $\theta=1$ and  $\tau_t T\equiv T$, which reduces to the results of \cite{gk1}.

The moments of the distribution function   (\ref{13a}) -- (\ref{13i}) 
reduce to
\ben\lb{13aa}
&&p_{ij}^{I}=\chi b\rho\left[\frac{5\kappa+2}{5(\kappa+1)}p_{\langle ij\rangle}+ nkT\left(1+\Delta\right)\delta_{ij}-\frac{\kappa\d}{\kappa+1}\sqrt{\frac{kT}{\pi m}}\rho\varepsilon_{ijk}s_k\right],
\\\lb{13ba}
&&q_i^I=\chi b\rho\left[\frac{5\kappa+3}{5(\kappa+1)}q_i-\frac{\kappa\d}{2(\kappa+1)}\sqrt{\frac{kT}{m\pi}}\varepsilon_{ijk}m_{jk}\right],\quad
h_i^I=\chi b \rho\left[\frac{2}{3(\kappa+1)}h_i-\frac{\kappa\d}{2(\kappa+1)}\sqrt{\frac{kT}{m\pi}}\varepsilon_{ijk}m_{jk}\right], 
\\\lb{13ca}
&&m_{ij}^I=0,\qquad p_{ijk}=\frac25(q_i\delta_{jk}+q_j\delta_{ik}+q_k\delta_{ij}),\qquad q_{ij}=\frac{kT}m\left[\frac{5}2nkT\left(1+2\Delta\right)\delta_{ij}+\frac72p_{\langle ij\rangle}\right],
\\\lb{13da}
&&p_{ijk}^I=\chi b\rho\bigg\{\frac{10\kappa+9}{25(\kappa+1)}\left[q_i\delta_{jk}+q_j\delta_{ik}+\frac{10\kappa+4}{10\kappa+9}q_k\delta_{ij}\right]-\frac{\kappa\d}{\kappa+1}\sqrt{\frac{kT}{m\pi}} m_{s(i}\varepsilon_{j)ks}\bigg\},
\\\lb{13ea}
&&q_{ij}^I=\chi b\rho\frac{kT}m\Bigg\{\left[\frac52+\frac{5\kappa^2+8\kappa+17}{(\kappa+1)^2}\Delta\right]nkT\delta_{ij}+\frac{35\kappa^2+54\kappa+17}{10(\kappa+1)^2}p_{\langle ij\rangle}-\frac{11}4\frac{\kappa\d}{\kappa+1}\sqrt{\frac{kT}{m\pi}}\rho\varepsilon_{ijk}s_k\Bigg\},
\\\no
&&m_{ijk}^I=\chi b\rho\left\{\frac{nkT}{5(\kappa+1)}\left[s_j\delta_{ik}+s_k\delta_{ij}-4s_i\delta_{jk}\right]+\frac{2nkT(\kappa-1)}{\d(\kappa+1)^2}\sqrt{\frac{kT}{\pi m}}\Delta\varepsilon_{ijk}\right.
\\\lb{13ga}
&&\left.\qquad+\frac2{5\d(\kappa+1)^2}\sqrt{\frac{kT}{\pi m}}\left[(6\kappa-2)\varepsilon_{ikr}p_{\langle jr\rangle}+(\kappa-5)\varepsilon_{ijr}p_{\langle kr\rangle}\right]\right\},
\;
m_{ijk}=0,\; h_{ij}=\frac{3kT}{2m}\left(nkT\delta_{ij}+p_{\langle ij\rangle}\right),
\\\lb{13ia}
&&
 h_{ij}^I=\chi b\rho\frac{kT}m\Bigg\{\left[\frac32+\frac{2\kappa}{(\kappa+1)^2}\Delta\right]nkT\delta_{ij}
+\frac{15\kappa^2+19\kappa+6}{10(\kappa+1)^2}p_{\langle ij\rangle}
-\frac{3\kappa\d}{2(\kappa+1)}\sqrt{\frac{kT}{\pi m}}\rho\varepsilon_{ijk}s_k\Bigg\},
\een
while the production terms (\ref{14c}) -- (\ref{14k}) become
\ben\lb{14ca}
&&\zeta=\zeta^I=0,\qquad
P_{ij}=-\frac83\nu\chi \Bigg\{nkT\frac{\kappa}{(1+\kappa)^2}\Delta\delta_{ij}+\frac{1}{20}\frac{6+13\kappa}{(1+\kappa)^2}p_{\langle ij\rangle}\Bigg\},
\\\no
&&P_{ij}^I=\chi b \rho\Bigg\{\frac6{5(\kappa+1)}nkT\frac{\partial v_{\langle i}}{\partial x_{j\rangle}}
+\frac{2(\kappa+3)}{25(\kappa+1)^2}\frac{\partial q_{(i}}{\partial x_{j)}}+\frac{2(3\kappa-1)}{25(\kappa+1)^2}\frac{\partial q_k}{\partial x_k}\delta_{ij}+\frac{4\kappa}{15(\kappa+1)^2}\bigg[\frac{\partial h_{(i}}{\partial x_{j)}}-2\frac{\partial h_k}{\partial x_k}\delta_{ij}\bigg]
\\\lb{14da}
&&\qquad+\frac\d5\sqrt{\frac{kT\tau_t}{\pi m}}\left[\frac{\kappa(6\kappa-2)}{(\kappa+1)^2}\frac{\partial m_{r(i}}{\partial x_k}\varepsilon_{j)kr}+\frac{\kappa(\kappa+3)}{(\kappa+1)^2}\frac{\partial m_{rs}}{\partial x_{(i}}\varepsilon_{j)sr}\right]\Bigg\},
\\\lb{14ea}
&&Q_i=-\frac{2}{15}\nu\chi\left\{\frac{17\kappa+4}{(\kappa+1)^2}q_i-\frac{25\kappa}{3(\kappa+1)^2}h_i\right\},
\\\lb{14fa}
&&Q_i^I=\chi b \rho\frac{kT}m\Bigg\{\frac{nk}{\kappa+1}\left(\frac{\partial T}{\partial x_i}+T\frac{\partial \Delta}{\partial x_i}\right)
+\frac{1}{10(\kappa+1)}\frac{\partial p_{\langle ij\rangle}}{\partial x_j}\Bigg\},
\\\lb{14ga}
&&M_i=-\frac4{3(\kappa+1)}\nu\chi\rho s_i,
\qquad
M_i^I=\chi b\rho\frac1{\kappa+1}\left\{\frac{2\nu\chi}{3\chi b\rho}\varepsilon_{ijk}\left[\rho\frac{\partial v_k}{\partial x_j}+\frac{m}{10k T\tau_t}\frac{\partial q_k}{\partial x_j}\right]+\frac35\frac{\partial m_{\langle ij\rangle}}{\partial x_j}+\frac{\partial m_{[ij]}}{\partial x_j}\right\},
\\\lb{14ha}
&&M_{ij}=\frac23\nu\chi\left[\frac{1}{3\d(\kappa+1)}\sqrt{\frac{m\pi}{kT}}\varepsilon_{ijk}h_k-2m_{ij}\right],
\\\lb{14ia}
&&M_{ij}^I=\chi b\rho\left\{nkT\frac{\partial s_i}{\partial x_j}-\frac{2}{\d(\kappa+1)}\sqrt{\frac{kT}{\pi m}}\left[2nk\varepsilon_{ijk}\frac{\partial T}{\partial x_k}+\varepsilon_{irs}\frac{\partial p_{\langle js\rangle}}{\partial x_r}\right]\right\},
\\\lb{14ja}
&&H_i=\nu\chi\Bigg\{\frac{2\kappa}{3(\kappa+1)^2}q_i-\frac{2(2\kappa^2+2\kappa+1)}{3(\kappa+1)^2}h_i
+\frac{\kappa\d}{2(\kappa+1)}\sqrt{\frac{\pi kT}{m}}\varepsilon_{ijk}m_{jk}\Bigg\},
\\\lb{14ka}
&&H_i^I=\chi b\rho \frac{k^2}{m^2}\rho T\frac{3\kappa+1}{2(\kappa+1)}\left\{\frac{\partial T}{\partial x_i}-T\frac{\partial \Delta}{\partial x_i}\right\}.
\een

The new constitutive equations (\ref{18a}) -- (\ref{18f}) read
\ben\lb{18aa}
&&p_{ij}^{II}=-\frac{4nkT}{\nu\chi}\frac{\chi^2b^2\rho^2}\pi\left[\frac{\partial v_k}{\partial x_k}\delta_{ij}+\frac{3\kappa}{2(\kappa+1)}\frac{\partial v_{[i}}{\partial x_{j]}}+\frac{21\kappa+12}{10(\kappa+1)}\frac{\partial v_{\langle i}}{\partial x_{j\rangle}}\right],\qquad \zeta^{II}=0,
\\\lb{18ba}
&&q_{i}^{II}=-6\frac1{\nu\chi}\frac{k^2}{m^2}\rho T\frac{\chi^2b^2\rho^2}\pi\frac{2\kappa+1}{\kappa+1}\frac{\partial T}{\partial x_i},
\qquad h_{i}^{II}=-6\frac1{\nu\chi}\frac{k^2}{m^2}\rho T\frac{\chi^2b^2\rho^2}\pi\frac1{\kappa+1}\frac{\partial T}{\partial x_i},
\\\lb{18da}
&&m_{ij}^{II}=\frac3{\nu\chi}nkT\frac{\chi^2b^2\rho^2}\pi\frac1{\kappa+1}\left[\frac35\frac{\partial s_{\langle i}}{\partial x_{j\rangle}}+\frac{\partial s_{[i}}{\partial x_{j]}}\right],
\qquad
M_i^{II}=\frac6{5\nu\chi}nkT\frac{\chi^2b^2\rho^2}\pi\frac1{\kappa+1}\left(\frac{\partial^2 s_j}{\partial x_i\partial x_j}-2\frac{\partial^2 s_i}{\partial x_j\partial x_j}\right).
\een

\section*{Acknowledments}  This work was supported by Conselho Nacional de Desenvolvimento Cient\'{i}fico e Tecnol\'{o}gico (CNPq), grant No.  304054/2019-4.


\begin{thebibliography}{0}
\bibitem{r1}  C. K. K. Lun, S. B. Savage, D. J. Jeffrey and N. Chepurniy, Kinetic theories for granular flow: Inelastic particles in Couette-flow and slightly inelastic particles in a general flowfield, J. Fluid Mech., 140 (1984), 223–256.
\bibitem{r2}  J. T. Jenkins and M. W. Richman, Grad’s 13-Moment system for a dense gas of inelastic spheres, Arch. Ration. Mech. Anal., 87 (1985), 355–377.
\bibitem{r3}  J. T. Jenkins and M. W. Richman, Kinetic theory for plane flows of a dense gas of identical, rough, inelastic, circular disks, Phys. Fluids, 28 (1985), 3485–3494.
\bibitem{r4} C.K.K. Lun, S.B. Savage, A simple kinetic theory for granular flow of rough, inelastic, spherical particles. J. Appl. Mech. 54, (1987) 47–53.
\bibitem{r5} A. Goldshtein and M. Shapiro, Mechanics of collisional motion of granular-materials. 1. General hydrodynamic equations, J. Fluid Mech., 282 (1995), 75–114.
\bibitem{r6}  J. J. Brey, J. W. Dufty, C. S. Kim and A. Santos, Hydrodynamics for granular flow at low density, Phys. Rev. E, 58 (1998), 4638–4653.
\bibitem{r7} V. Garz\'o and J. W. Dufty, Dense fluid transport for inelastic hard spheres,Phys. Rev. E 59,  (1999) 5895–5911.
\bibitem{r8} D. Risso and P. Cordero, Dynamics of rarefied granular gases Phys. Rev E 65, (2002) 021304–9.
\bibitem{r9} I. Goldhirsch, Rapid granular flows, Ann. Rev. Fluid Mech 35, (2003) 267–293.

\bibitem{r10} M. Bisi, G. Spiga, and G. Toscani, Grad’s equations and hydrodynamics for weakly inelastic granular flows Phys. Fluids 16, (2004) 4235–4247.
\bibitem{r11} V. Garz\'o, Instabilities in a free granular fluid described by the Enskog equation Phys. Rev. E 72, (2005) 021106–9.
\bibitem{r12} A. Santos, G.M. Kremer, V. Garz\'o, Energy production rates in fluid mixtures of inelastic rough hard spheres. Prog. Theor. Phys. Suppl. 184,  (2010) 31–48.
\bibitem{r13} G. M. Kremer and W. Marques-Jr., Fourteen moment theory for granular gases, Kinet. Relat. Mod. 4,  (2011) 317–331.
\bibitem{r14} A. Santos, G. M. Kremer and M. dos Santos, Sonine approximation for collisional moments of granular gases of inelastic rough spheres. Phys. Fluids 23,  (2011) 030604. 
\bibitem{r15} V. Garz\'o, Grad's moment method for a granular fluid at moderate densities Phys. Fluids 25,  (2013) 043301–22.
\bibitem{r16} G.M. Kremer, A. Santos, V. Garz\'o, Transport coefficients of a granular gas of inelastic rough hard spheres. Phys. Rev. E 90,  (2014) 022205.

\bibitem{r17} V. Garz\'o, A. Santos, G.M. Kremer, Impact of roughness on the instability of a free-cooling granular gas. Phys. Rev. E 97,  (2018) 052901.
\bibitem{r18} V. K. Gupta, P. Shukla, and M. Torrilhon, Higher-order moment theories for dilute granular gases of smooth hard spheres J. Fluid Mech. 836,  (2018) 451–501.
\bibitem{r19} G. M. Kremer, Fourteen Moment Method for Moderately Dense Granular Gases, 31st International Symposium on Rarefied Gas Dynamics
AIP Conf. Proc. {\bf2132}, (2019) 080002.
\bibitem{r20} G.M. Kremer, A. Santos, Granular gas of inelastic and rough Maxwell particles. J. Stat. Phys. 189,  (2022) 23. 

\bibitem{r21} V. Garz\'o, Towards a better understanding of granular
flows, J. Fluid Mech. 968, (2023) F1. 

\bibitem{b1}N. V. Brilliantov and T. P\"oschel, \emph{Kinetic Theory of Granular Gases} (Oxford University Press, Oxford, 2004).

\bibitem{b2} V. Garz\'o, \emph{Granular Gaseous Flows. A Kinetic Theory Approach to Granular Gaseous Flows} (Springer, Cham, 2019)

\bibitem{r22} F.B. Pidduck, The kinetic theory of a special type of rigid molecule. Proc. R. Soc. Lond. A 101,  (1922) 101–112.

\bibitem{CC} S. Chapman and T. G. Cowling, \emph{The Mathematical Theory of Non-Uniform Gases} 3rd edn. (Cambridge University Press, Cambridge, 1970).

\bibitem{gk0} G. M. Kremer, \emph{An Introduction to the Boltzmann Equation and Transport Processes in Gases} (Springer, Berlin, 2010).

\bibitem{r23} B.J. McCoy, S.I. Sandler, J.S. Dahler, Transport properties of polyatomic fluids. IV. The kinetic theory of a dense gas of perfectly rough spheres. J. Chem. Phys. 45(10), (1966) 3485–3512. 

\bibitem{gk1} D. C. Gaio and G. M. Kremer, Kinetic theory for polyatomic dense gases of rough spherical molecules, J. Non-Equilib. Thermodyn. {16} (1991) 357-379.

\bibitem{Lun} C. K. K. Lun, Kinetic theory for granular flow of dense, slightly inelastic, slightly rough spheres, J . Fluid Mech. (1991), 233, 
539-559.

\bibitem{r30} A. Santos and G. M. Kremer, Exact transport coefficients from the inelastic rough Maxwell model of a granular gas, arXiv: 2311.15440. 

\bibitem{Ree}  F. H. Ree and  W. G. Hoover, Seventh virial coefficients for hard spheres and hard disks, J. Chem. Phys. {46}, (1967) 4181–4197.

\bibitem{gk2} G. M. Kremer and C. Cercignani, Wave Propagation in a Rarefied Gas of Rough Spheres, in \emph{Proceedings of the 15th International Symposium on Rarefied Gas Dinamics}, edited by  V. Boffi and C. Cercignani (B. G. Teubner, Stuttgart, 1986) pp. 110-121.

\bibitem{fr1} N. V. Brilliantov, T. P\"oschel, W. T. Kranz, and A. Zippelius, Translations and rotations are correlated in granular gases, PRL 98 (2007), 128001.
\bibitem{fr2} B. Gayen and M. Alam, Orientational correlation and velocity distributions in uniform shear flow
of a dilute granular gas, PRL 100 (2008), 068002.
\bibitem{fr3} B. Gayen and M. Alam, Effect of Coulomb friction on orientational correlation and velocity distribution functions
in a sheared dilute granular gas, Phys. Rev. E 84 (2011), 021304.

\bibitem{fr4} R. Rongali and M. Alam, Higher-order effects on orientational correlation and relaxation dynamics in homogeneous cooling
of a rough granular gas, Phys. Rev. E 89 (2014), 062201. 

\bibitem{Cow} S. C. Cowing, The theory of polar fluids, Adv. Appl. Mech., 14 (1974) 279-347. 

\bibitem{Les} F. M. Leslie, On the thermodynamics of polar fluids, Arch. Rational Mech. Anal., 70 (1979), 189-202.





\end{thebibliography}
\end{document}